\newcommand{\msol}{\mbox{\,M$_\odot$}}
\newcommand{\mdotrate}{\mbox{\,M$_\odot$\,yr$^{-1}$}}
\newcommand{\kms}{\mbox{\,km\,s$^{-1}$}}
\documentclass[a4,useAMS,usenatbib]{mnras}

\usepackage{graphicx,subfigure}

\title[Massive star formation simulations]{Radiation-hydrodynamical simulations of massive star formation using Monte Carlo radiative transfer \\ II. The formation of a 25 solar-mass star}
\author[Tim J. Harries et al.]{Tim J. Harries\thanks{E-mail:
th@astro.ex.ac.uk}, Tom A. Douglas, and Ahmad Ali \\
Department of Physics and Astronomy, University of Exeter, Stocker Road, Exeter EX4 4QL, United Kingdom\\}
\begin{document}

\date{}

\pagerange{\pageref{firstpage}--\pageref{lastpage}} \pubyear{2016}

\maketitle

\label{firstpage}

\begin{abstract}
We present a numerical simulation of the formation of a massive star using Monte-Carlo-based radiation hydrodynamics (RHD). The star forms via stochastic disc accretion and produces fast, radiation-driven bipolar cavities. We find that the evolution of the infall rate (considered to be the mass flux across a 1500\,au spherical boundary), and the accretion rate onto the protostar, are broadly consistent with observational constraints. After 35\,kyr the star has a mass of 25\msol\ and is surrounded by a disc of mass 7\,\msol\ and 1500\,au radius, and we find that the velocity field of the disc is close to Keplerian. Once again these results are consistent with those from recent high-resolution studies of discs around forming massive stars. Synthetic imaging of the RHD model shows good agreement with observations in the near- and far-IR, but may be in conflict with observations that suggests that MYSOs are typically circularly symmetric on the sky at 24.5\,\micron. Molecular line simulations of a CH$_3$CN transition compare well with observations in terms of surface brightness and line width, and indicate that it should be possible to reliably extract the protostellar mass from such observations.

\end{abstract}


\section{Introduction}

Although the paradigm for low-mass star formation has achieved a broad consensus, the mechanism by which high mass stars form is rather less clear (e.g. \citealt{tan_2014}). Observationally this is because the rapidity and rarity of massive star formation means that the process occurs while the stars are deeply embedded and that we observe it at great distances. However rapid progress is now being made, with the advent of high resolution observations of massive young stellar objects (MYSOs) using ALMA, that encode geometrical, thermal, chemical and dynamical information on the protostars, discs, and envelopes--for a recent review see \cite{beltran2016}. 

From a theoretical perspective massive star formation is challenging because theorists must deal not only with the hydrodynamics of self-gravitating collapse, but also with the enormous radiation field produced by the massive protostar, which not only heats and ionises the gas, but also results in forces that act to resist the collapse, leading to outflows and disc winds.   Progress in numerical simulations has followed a pattern of increasing complexity in the treatment of these radiative-feedback processes. The first calculations relied the diffusion approximation, which is easy to implement and computationally cheap and captures the physics of the radiation field well in the optically thick limit. The diffusion approximation breaks down as the radiation-field becomes free-streaming, and a flux-limiter is used in optically thin regions to overcome this drawback. Flux-limited diffusion (FLD) has been used successfully to model high-mass star formation  \citep{yorke_2002,krumholz_2009}.

One potential pitfall of using grey FLD to study high-mass star formation is the enormous disparity between the shape of the radiation field of the protostellar sources (often characteristic of temperatures exceeding 10\,kK) and that of the thermal dust emission ($<1500$\,K). The fact that dust absorption and scattering cross sections are much larger in the blue means that grey FLD may significantly underestimate the radiation forces from the direct (unprocessed) protostellar radiation field. This drawback is avoided by the so-called hybrid RT scheme, in which a frequency-dependent ray-tracing integral of the RT equation from the protostellar source is used to determine the intensity of the radiation field until it becomes optically thick, and the radiation field at this point is used as a boundary condition for the diffusion approximation in the rest of the computational domain. This method was first used in 1-D simulations of massive star formation by \cite{wolfire1986,wolfire1987} and was extended to higher dimensions by \cite{kuiper_2010a}. Although  more computationally expensive than FLD, the method has the advantage that it captures the physics of the first absorption of the protostellar radiation field by the dust, as well as any shadowing effects close to the source. Tests of the method show that it does a remarkably good job of determining the dust temperature when compared to dedicated RT codes \citep{kuiper_2013}. Many contemporary studies now employ this hybrid method \citep{kuiper_2012, klassen_2016, rosen_2016}.

Of course one would really like to conduct the RT step of the RHD simulation using a method that treats both the dust and gas opacities and emissivities in their full polychromatic glory. Dedicated RT codes such as {\sc cloudy} \citep{ferland_2013} manage this, as do a large number of Monte Carlo (MC) based codes \citep{bjorkman_2001, harries_2004, wood_2004,ercolano_2005, pinte_2006, robitaille_2011}, but these are usually perceived to be too slow to be useful as the radiation step in an RHD code. However the MC method has one distinct advantage: photon packets are independent. This means that MC loops are embarrassingly parallelisable, and one can use computer science to exploit the full power of modern distributed supercomputers to bring down the overhead of a MC RT step to something approaching that of the hydrodynamics step itself \citep{haworth_2012a}.  Furthermore once the simulation is complete the temperature of the dust and gas, and the photoionisation state of the gas may be used used to post-process the simulation and create synthetic observables, allowing a direct comparison with observations \citep{haworth_2012b}.

In \cite{harries_2015} we  presented a new MC method for including radiation pressure in RHD simulations that incorporates a level of microphysical detail that is significantly greater than that of flux-limited diffusion or hybrid techniques. We have shown that the new method works in both the pure absorption and pure scattering regimes, and properly treats anisotropic scattering processes. The method comprises a simple addition to the MC estimators required for radiative and photoionisation equilibrium calculations, and does not therefore represent a substantial computational overhead to the MC RHD methods described by \cite{haworth_2012a}. 

In this second paper of the series we apply the new method to the follow the formation of a massive star in three dimensions. In Section~\ref{method_sec} we review our numerical method and in Section~\ref{init_sec} we describe our initial conditions. In Section~\ref{model_sec} we describe the evolution of the model, and in Section~\ref{obs_sec} we compare our simulation with observational constraints. We discuss our model in the context of recent observations in Section~\ref{discussion_sec} and we make our conclusions and discuss future directions of our research in Section~\ref{conc_sec}. 

\section{Numerical method}
\label{method_sec}

We employ the {\sc torus} radiation hydrodynamics code \citep{harries_2000, harries_2004, harries_2011, harries_2015} which uses a MC based treatment of the radiation field coupled with an adaptive mesh hydrodynamics scheme and a Lagrangian sink particle implementation. We direct the reader to the first paper in this series \citep{harries_2015} for a detailed description of the numerical methods used to solve the RHD equations. We emphasise that we have extensively tested and benchmarked the MC RT scheme against dust continuum benchmarks \citep{harries_2004, pinte_2009} and H\,{\sc ii} region models \citep{haworth_2012a, bisbas_2015} as well as time-dependent heat-transfer benchmarks \cite{harries_2011}. We have also tested the coupled radiation-hydrodynamics scheme against the usual 1-D benchmarks, such as the Sod shock tube, the expansion of an H\,{\sc ii} region \citep{harries_2015, haworth_2015}, and radiation-driven shell tests \citep{harries_2015}. The Lagrangian sink particle scheme has been benchmarked against standard n-body tests as well as `standard' star formation tests such as Bondi accretion, Shu collapse, and Bondi-Hoyle accretion models \citep{harries_2015}.

\section{Initial conditions}
\label{init_sec}

Our calculation starts with a 100\,\msol\ cloud with a 0.1\,pc radius and an  $\rho(r) \propto r^{-2}$ density gradient. We impose solid body rotation with an angular frequency of $5 \times 10^{-13}$\, s$^{-1}$. We assume  a $10^{-12}$\,g\,cc$^{-1}$ threshold density of accretion onto the sink particle (i.e. mass above this threshold within the sink accretion radius is assumed to accrete onto the sink). We start with a 0.001\,\msol\ sink particle, which is held fixed at the origin. Our computational domain is an $8 \times 10^{17}$\,cm-sided cube, with a minimum of 7 and maximum of 12 levels of refinement.  The grid is refined in nested spherical volumes, with the inner 150\,au refined at 13\,au resolution, then 26.1\,au resolution beyond this out to 300\,au, 52.2\,au out to 500\,au, 104\,au out to 1000\,au, 209\,au to 2100\,au, and 418\,au for the rest of the domain.  For the purposes of this initial calculation the grid is not refined or unrefined as the calculation proceeds. We impose outflow/no inflow  conditions at all boundaries.

\subsection{Dust model}

We assume the dust is composed of silicate grains \citep{draine_1984}
in an MRN \citep{mathis_1977} size distribution ($n(a) \propto
a^{-3.5}$) with a minimum size of 0.005\,$\mu$m and a maximum size of
0.25\,$\mu$m. For the purposes of this calculation we keep this size
distribution constant in time and space, although we note that
detailed calculations suggest that coagulation and shattering may lead
to significant modification of the dust size distribution during
collapse \citep{suttner_1999}. We further assume that the gas and
  dust are strongly dynamically coupled, and that the dust sublimates
  when its temperature exceeds 1500\,K.

\subsection{The protostar model}

We take the mass of the sink particle and interpolate in the $\log(\dot{M})=-3$ evolutionary models of \cite{hosokawa_2009} to obtain the luminosity and radius of the protostar. We use the sink particle accretion rate to calculate the accretion luminosity (using the \cite{hosokawa_2009} radius) and assume that all the accreted kinetic power is converted to protostellar luminosity. The spectral energy distribution is found by interpolating in a grid of Kurucz model atmospheres using the temperature and surface gravity of the protostar, and adding on a blackbody contribution for the accretion luminosity with a temperature derived by assuming the accretion luminosity is liberated uniformly across the protostellar surface.

\section{Results}
\label{model_sec}

\begin{figure}
\includegraphics[width=80mm]{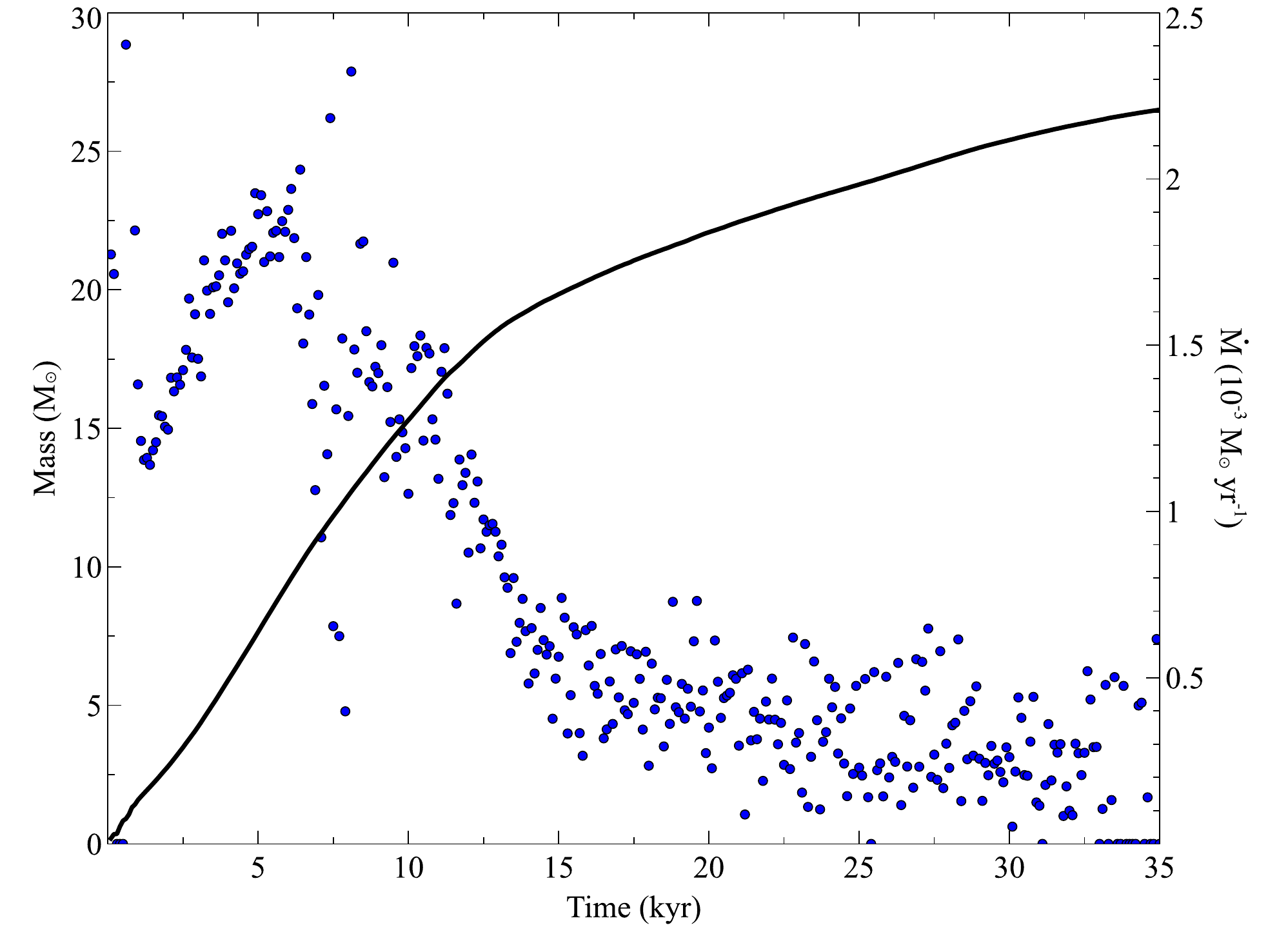}
\caption{The protostellar mass in our simulation plotted as a function of time (black line, left-hand ordinate). Also plotted is the instantaneous accretion rate onto the protostar (blue circles, right-hand ordinate).  }
\label{accretion_fig}
\end{figure}

The accretion rate at early times is about $10^{-3}$\mdotrate, rising
to a maximum of $2\times 10^{-3}$\mdotrate after 6\,kyr when the
accretion rate starts to decline as the disc forms and effects of
radiation feedback start to grow (the protostellar mass has reached
about 10\msol\ at this point, see Figure~\ref{accretion_fig}). Note
  that a non-rotating model with our mass and density profile that is
  ballistically infalling would have an accretion rate of $2 \times
  10^{-3}$\,\msol\,yr$^{-1}$ \citep{whitworth2001}. We plot the
  instantaneous accretion rate in Figure~\ref{accretion_fig}, which is
  the accretion rate at the time the code dumps the hydrodynamical
  data (every 100\,yrs). The accretion rate show significant
  stochastic variability throughout the run, at early times due to the
  feedback between increases in accretion luminosity (which dominates
  at this stage) and growth in thermal pressure. The protostellar
  growth is characterised by stochastic disc accretion at later
  times.

\begin{figure*}
\begin{center}$
\begin{array}{ccc}
             \includegraphics[width=0.33\textwidth]{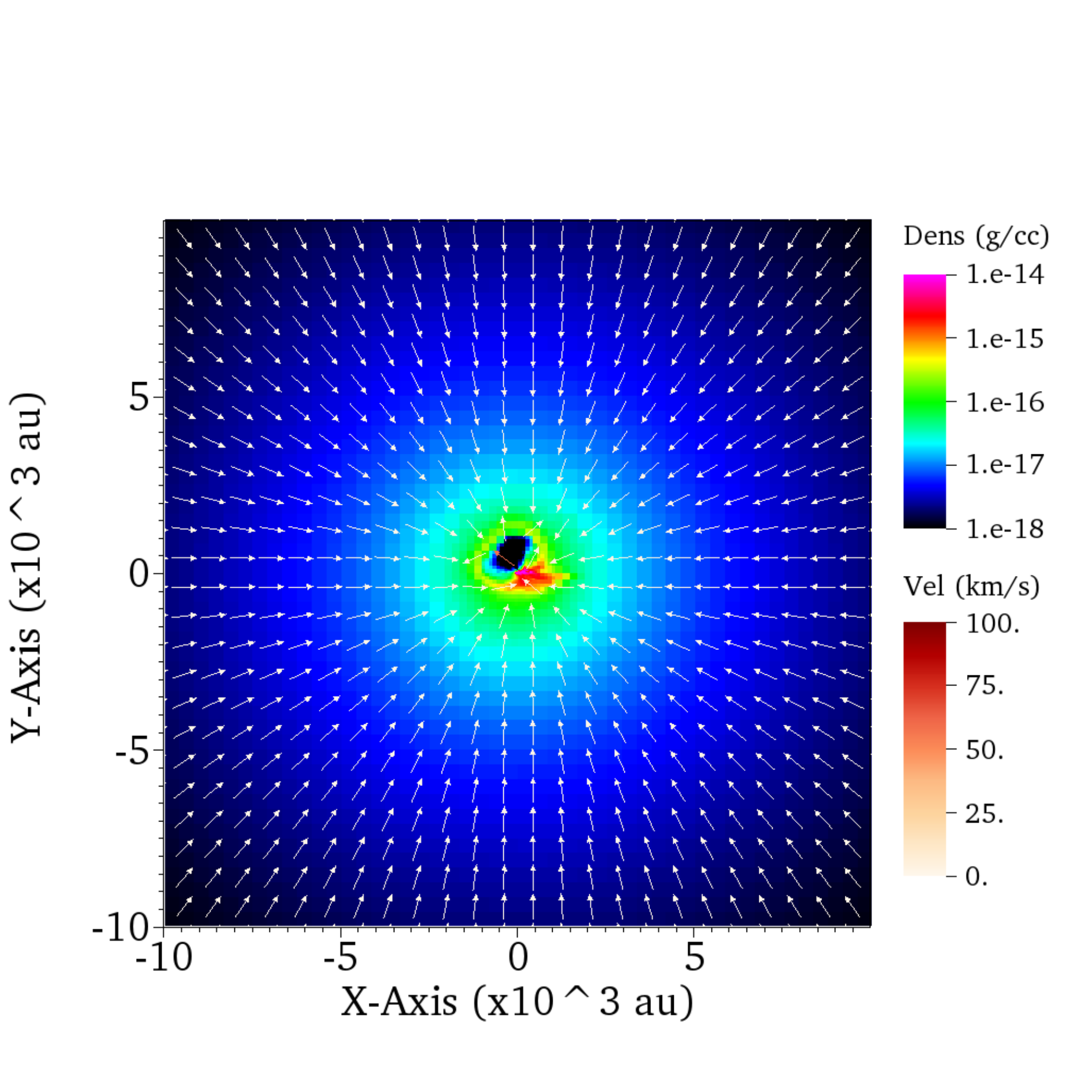}
             \includegraphics[width=0.33\textwidth]{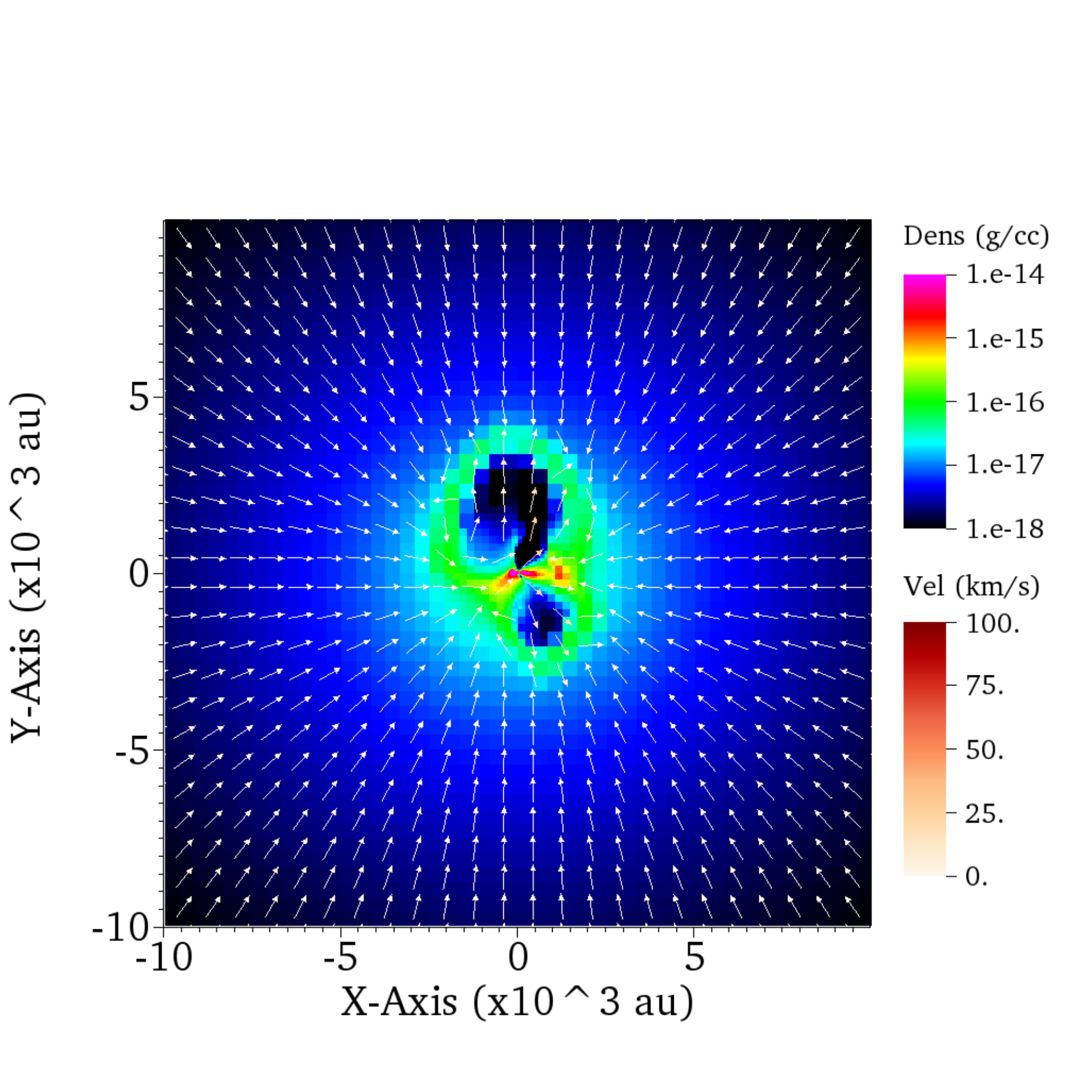}
            \includegraphics[width=0.33\textwidth]{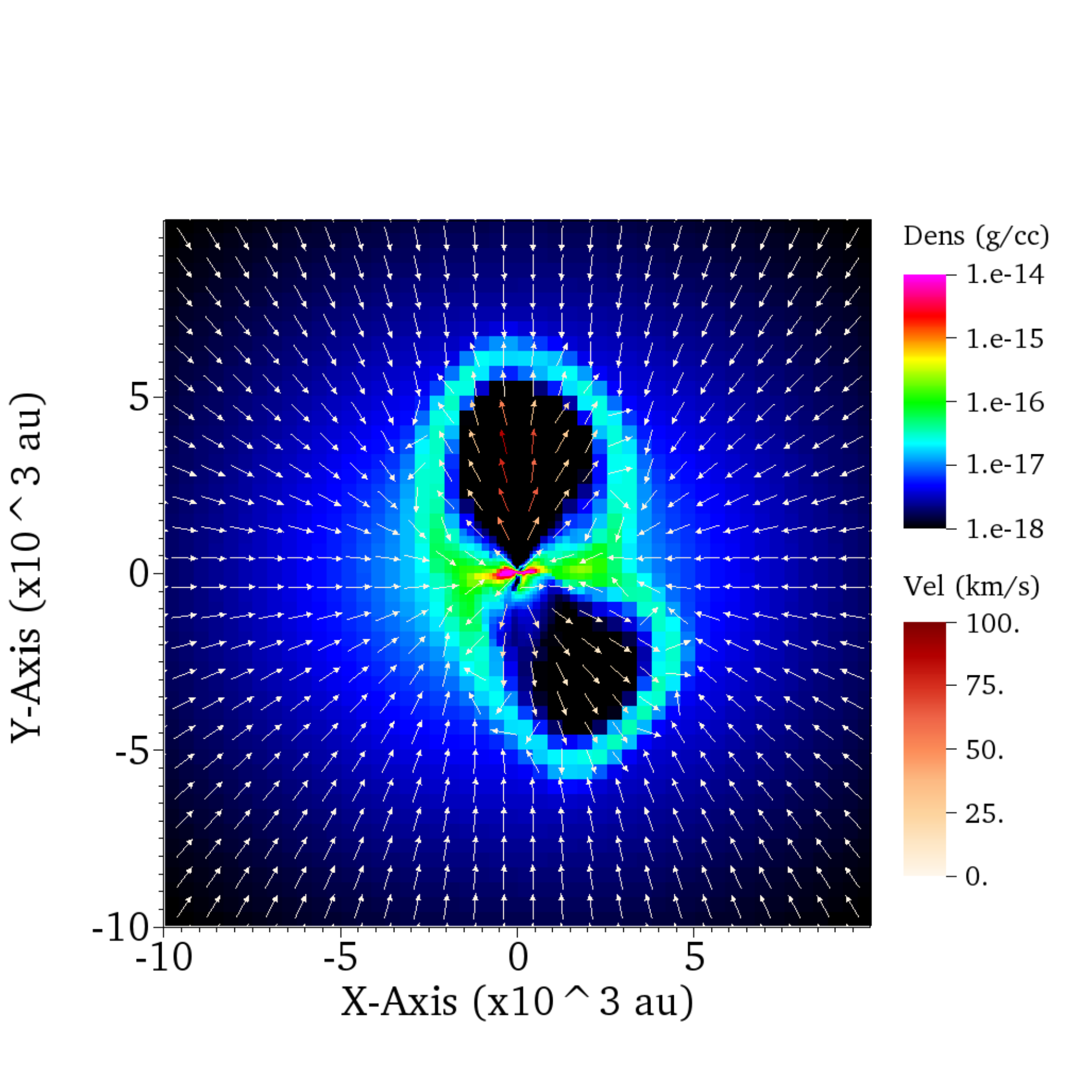} \\
             \includegraphics[width=0.33\textwidth]{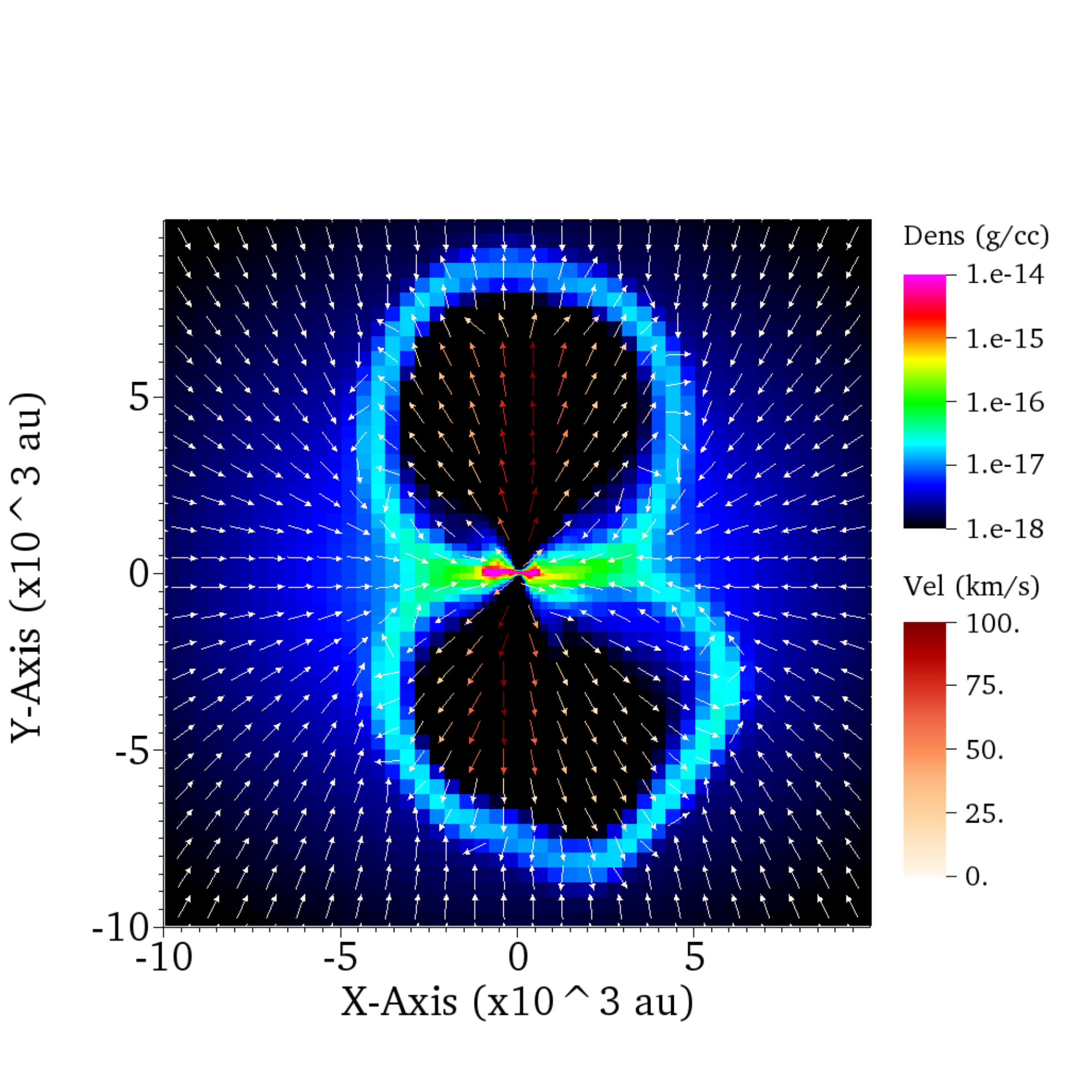}
            \includegraphics[width=0.33\textwidth]{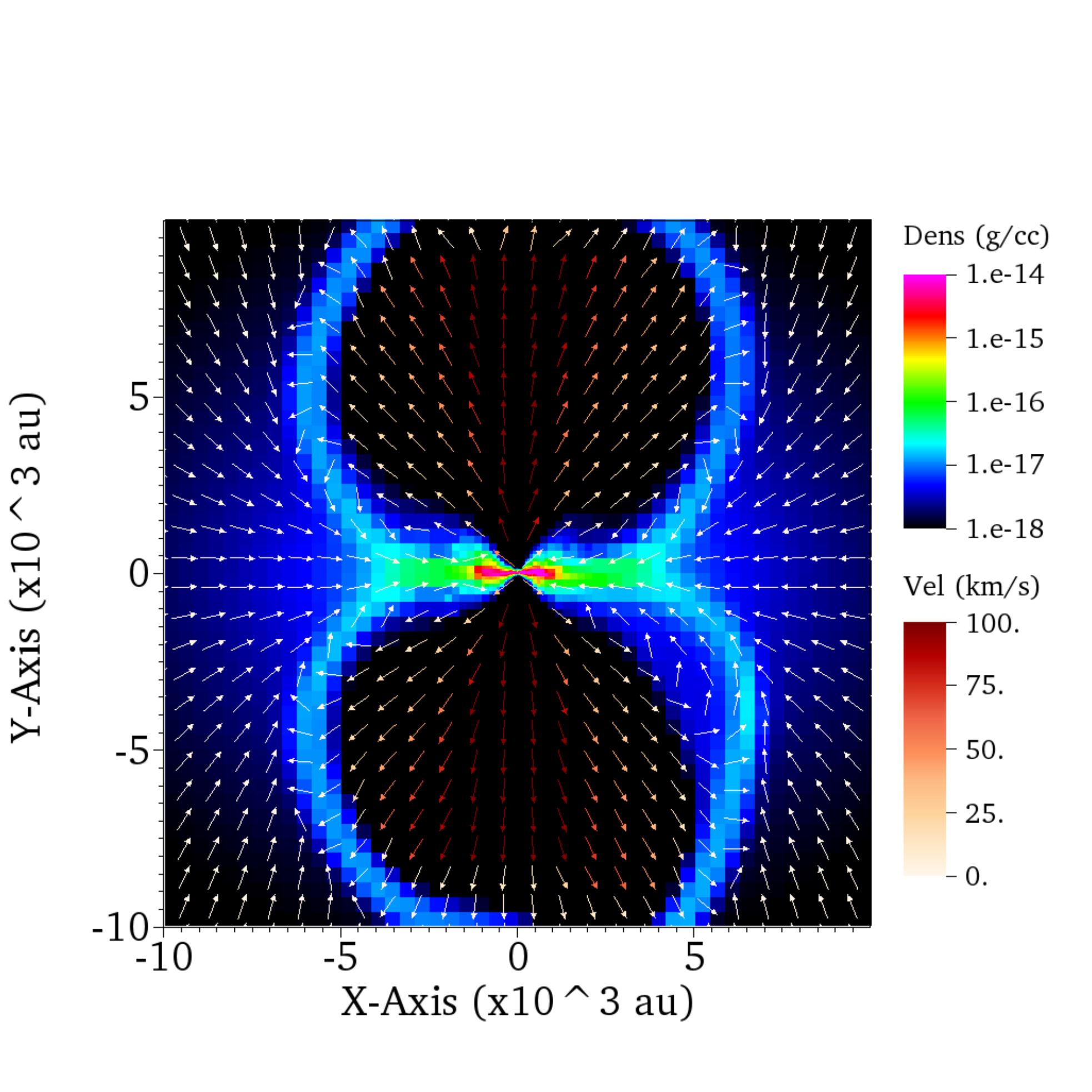}
             \includegraphics[width=0.33\textwidth]{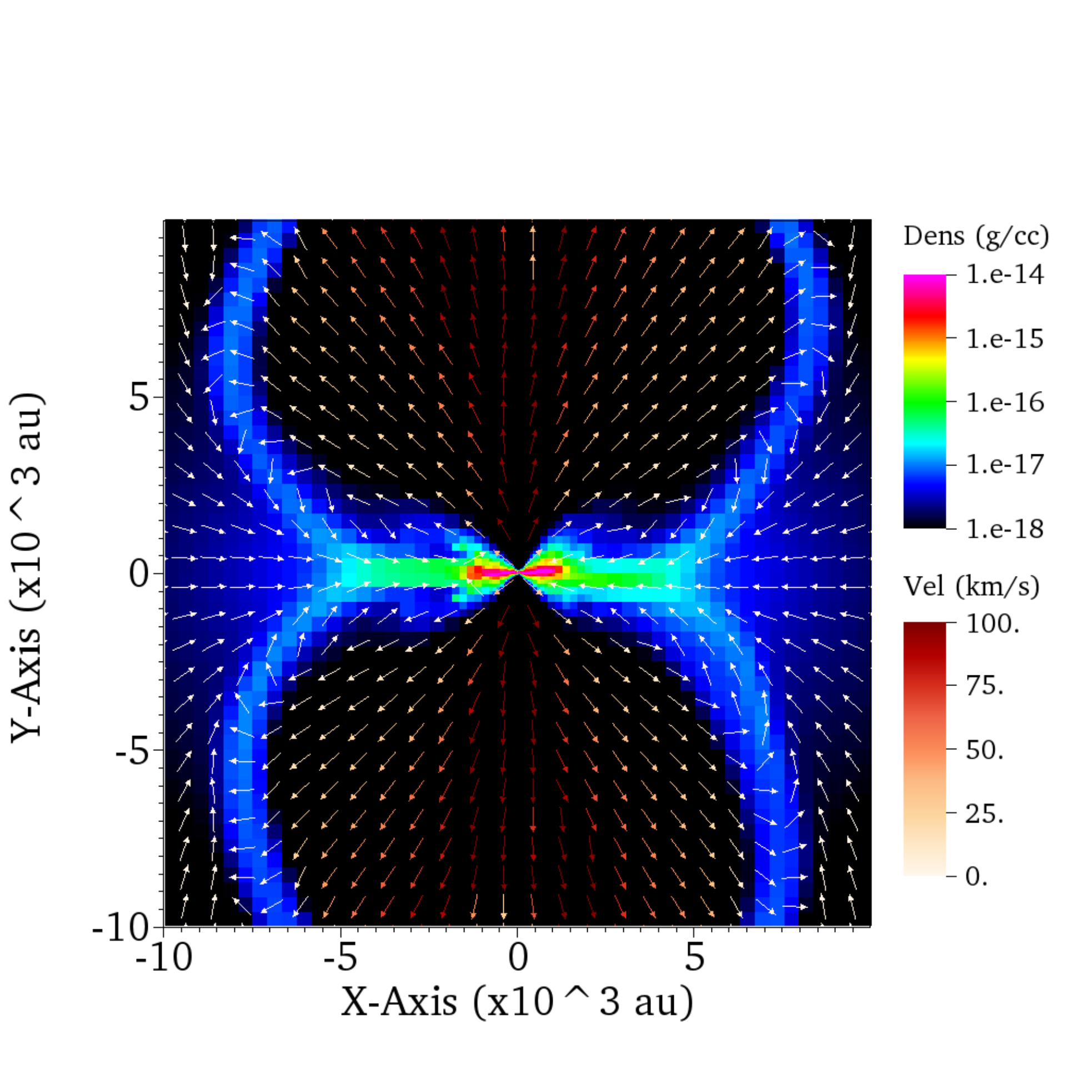} 
            \end{array}$
\end{center}
\caption{Density cuts through the $y$-axis of the simulation. The colour scale shows density (scaled between $10^{-19}$\,g\,cc$^{-1}$ and 10$^{-14}$\,g\,cc$^{-1}$). The arrows show the direction of the velocity field, and their colour the speed (between 0 and 100\,km\,s$^{-1}$.). The top three panels are at (left to right) 10\,kyr, 15\,kyr, and 20\,kyr and the bottom three panels are 25\,kyr, 30\,kyr, and 35\,kyr.}
\label{slice_y_fig}
\end{figure*}

\begin{figure*}
\begin{center}$
\begin{array}{ccc}
            \includegraphics[width=0.33\textwidth]{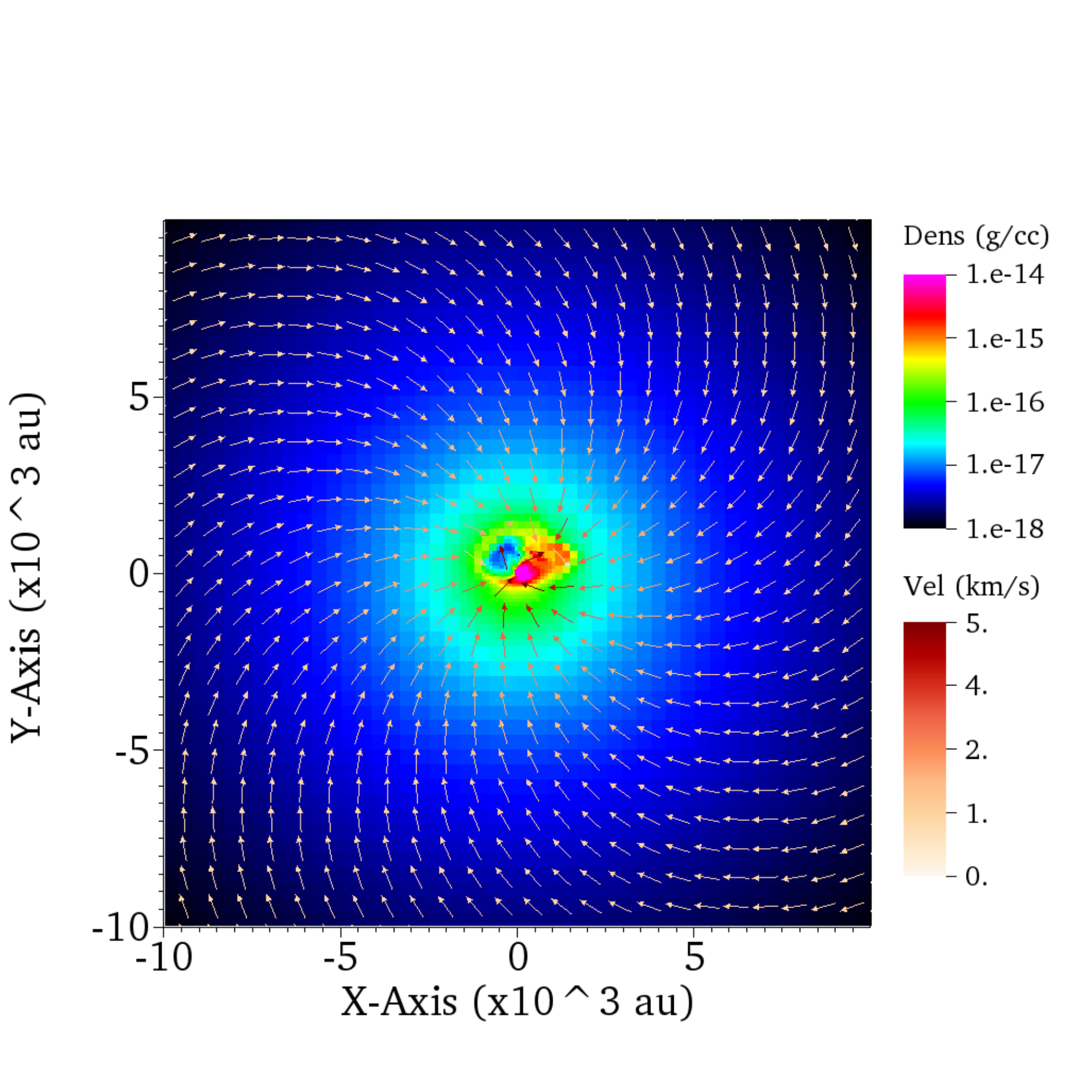}
            \includegraphics[width=0.33\textwidth]{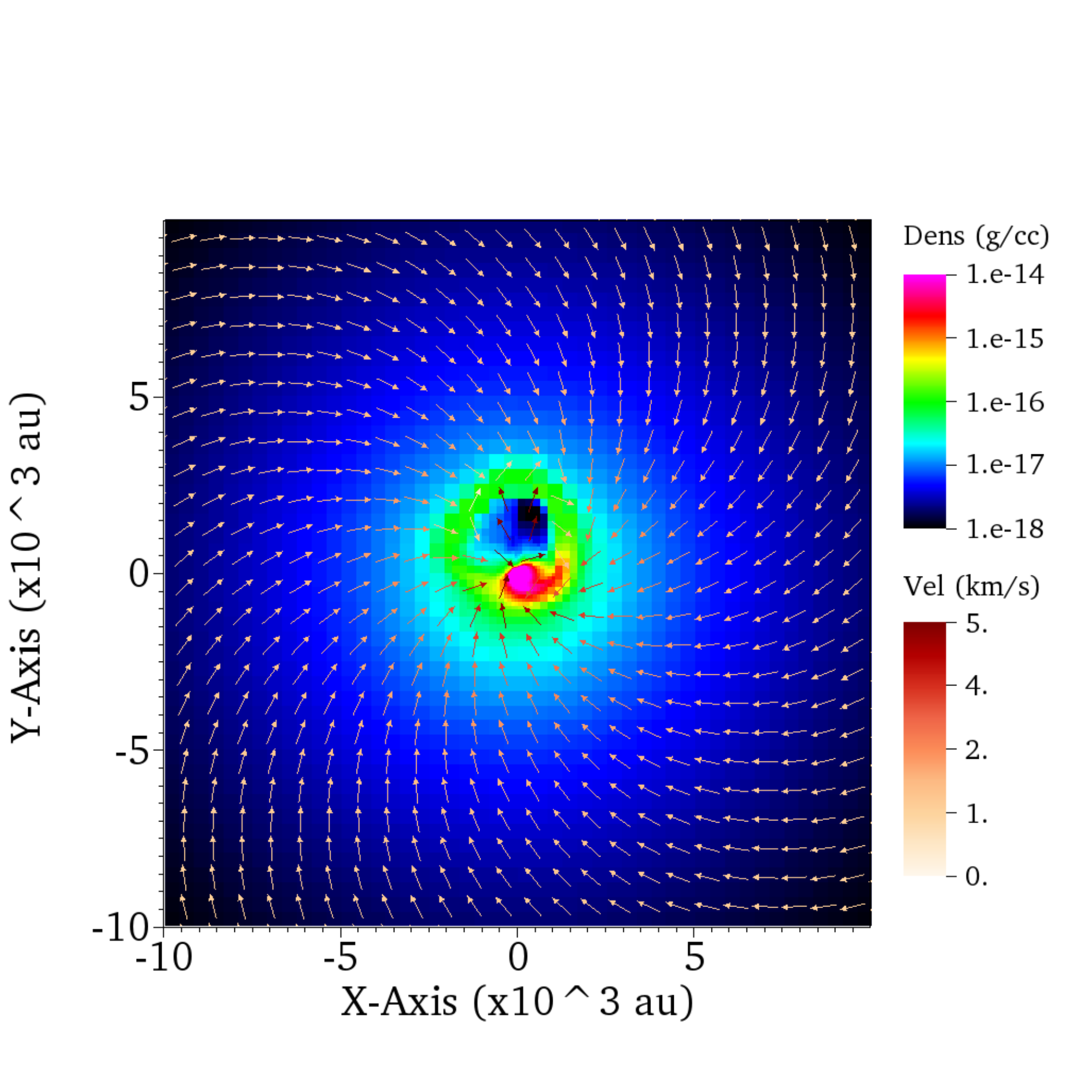}
             \includegraphics[width=0.33\textwidth]{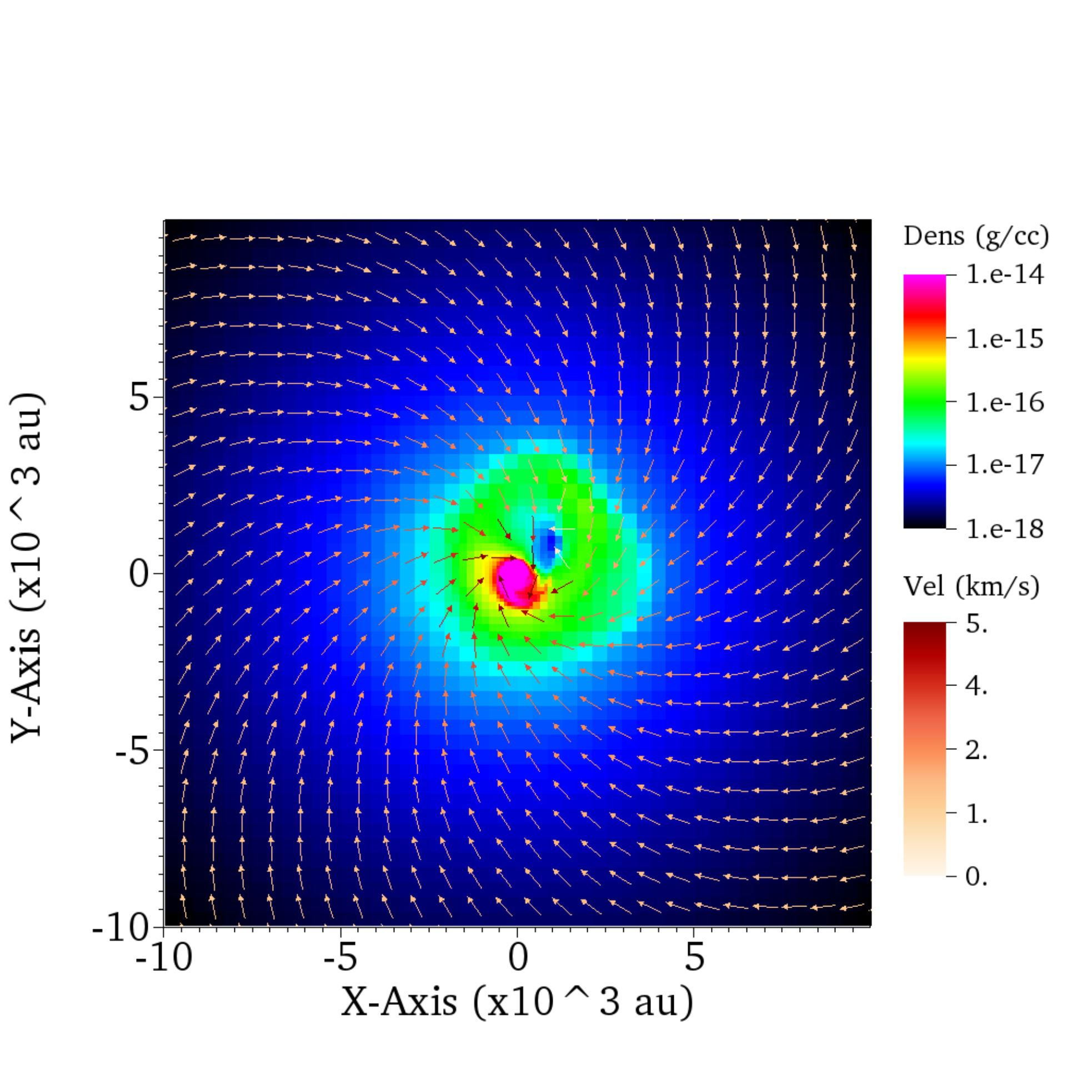} \\
              \includegraphics[width=0.33\textwidth]{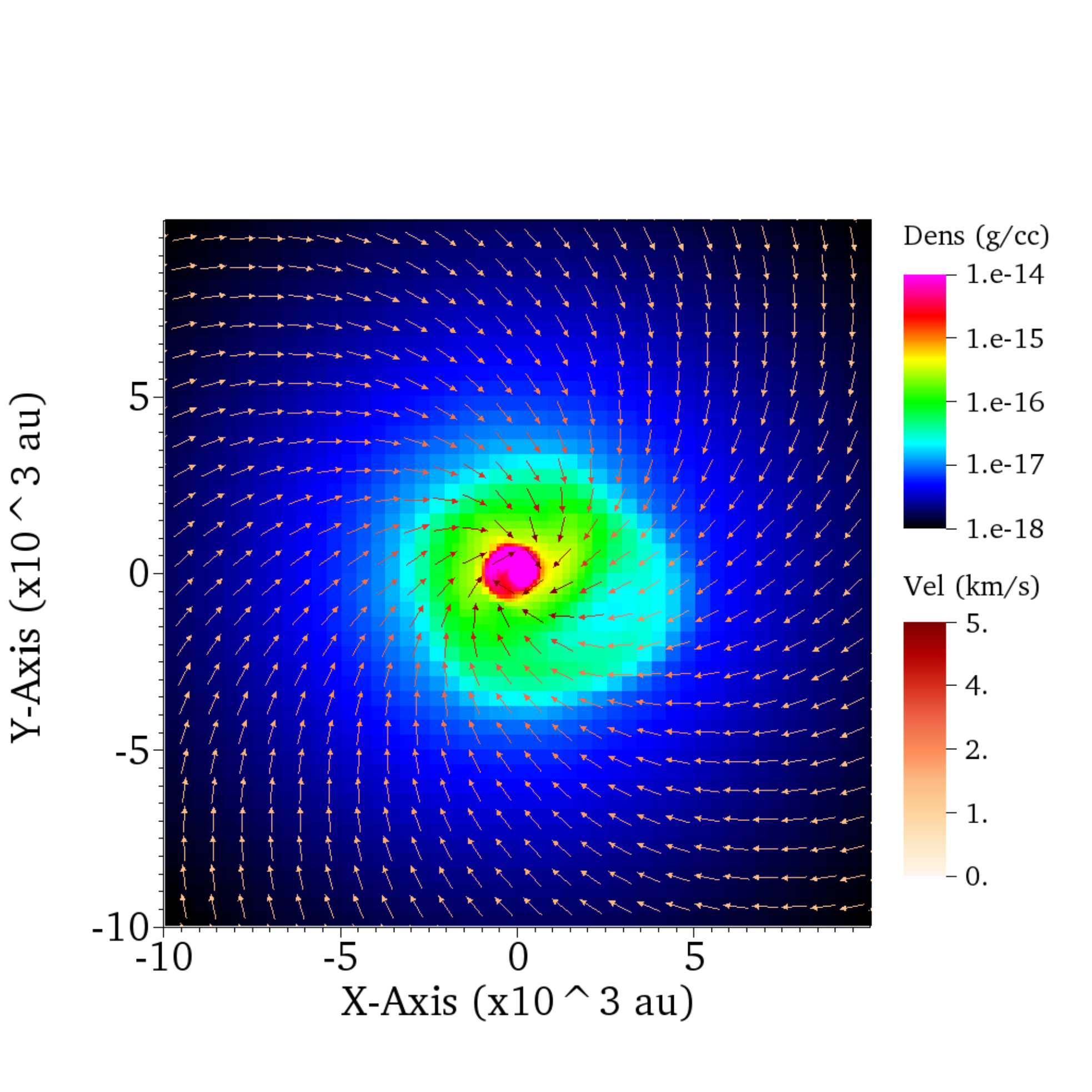}
              \includegraphics[width=0.33\textwidth]{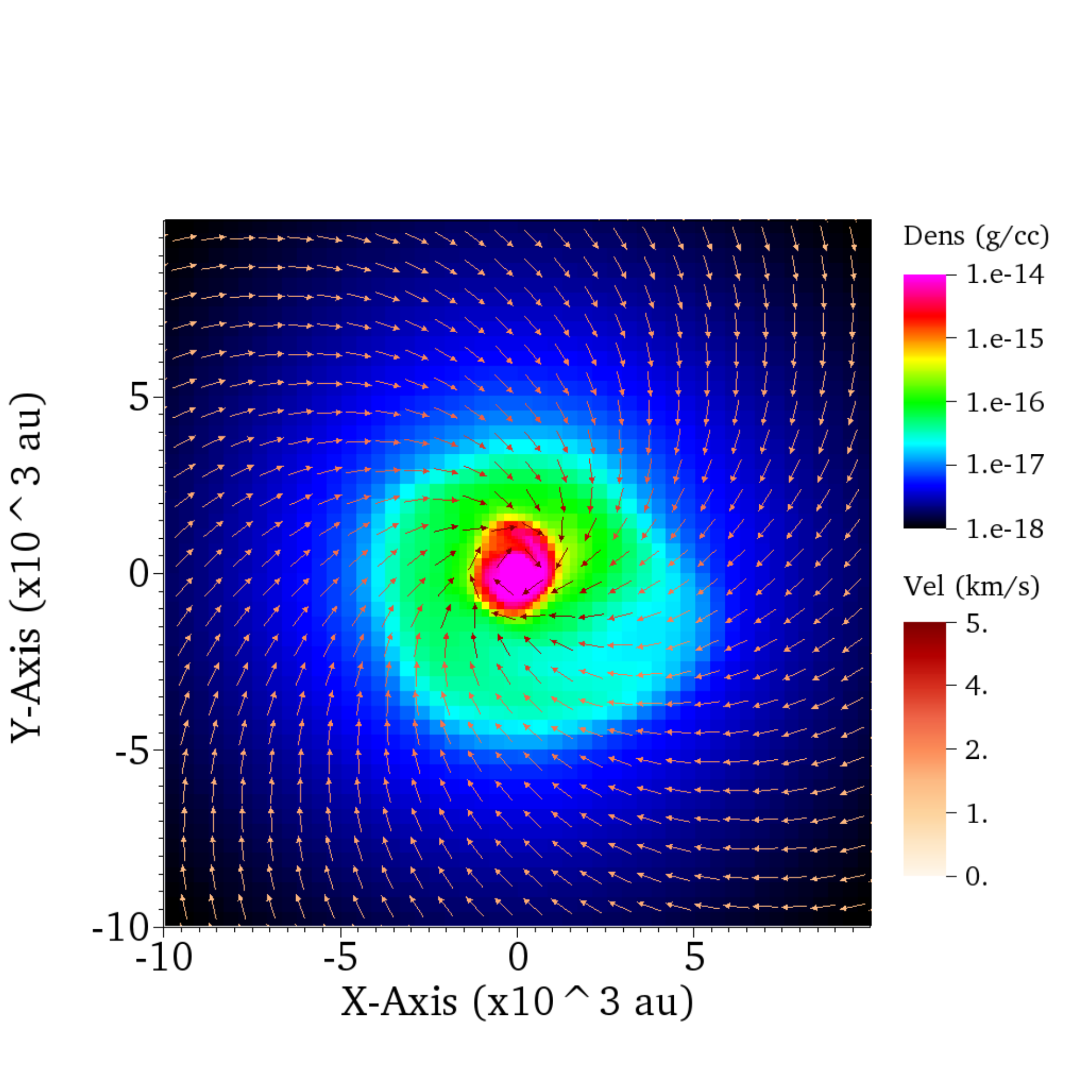}
              \includegraphics[width=0.33\textwidth]{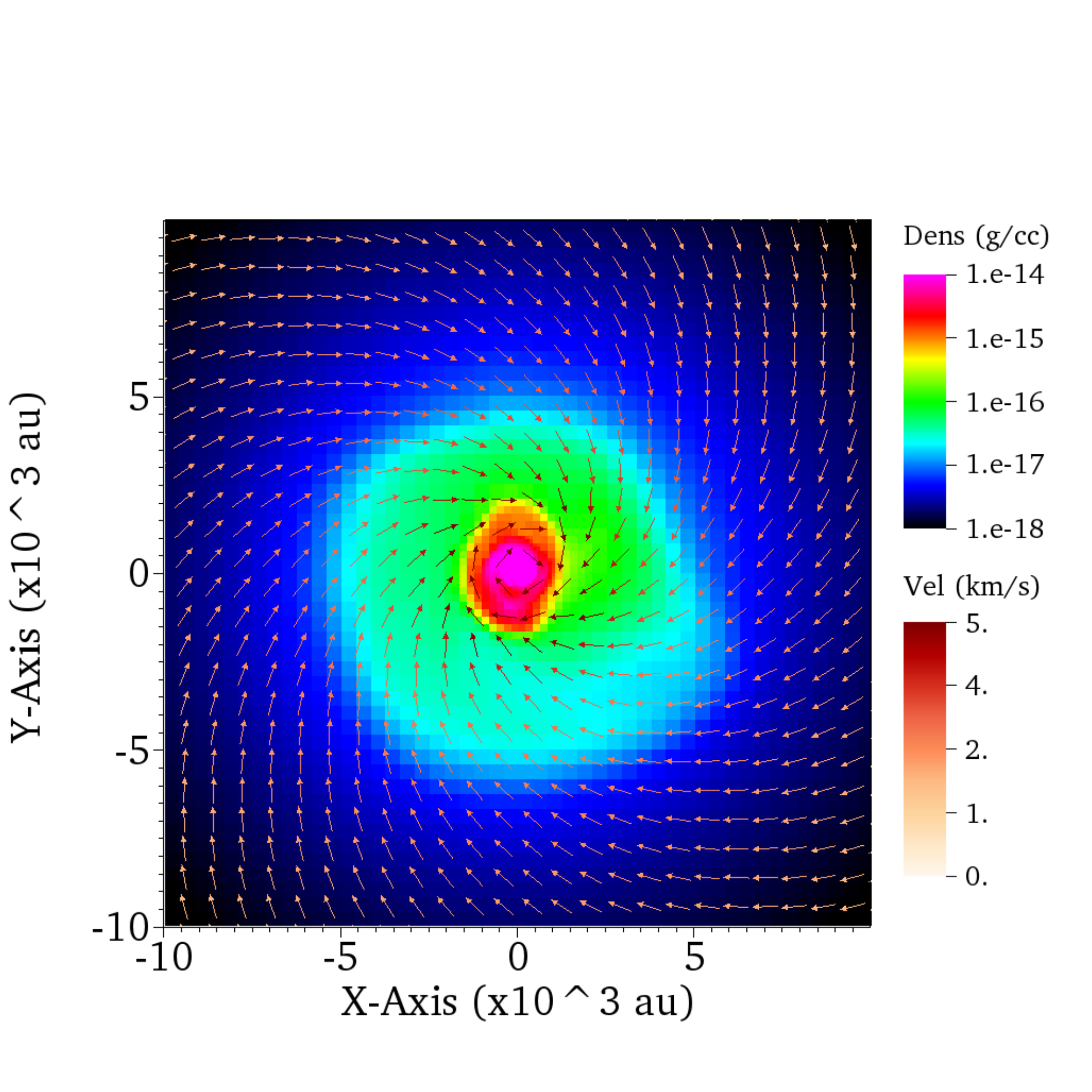}
            \end{array}$
\end{center}
\caption{Density cuts through the $z$-axis of the simulation. The colour scale show density (scaled between $10^{-19}$\,g\,cc$^{-1}$ and 10$^{-14}$\,g\,cc$^{-1}$). The arrows show the direction of the velocity field, and their colour the speed (between 0 and 5\,km\,s$^{-1}$.). The top three panels are at (left to right) 10\,kyr, 15\,kyr, and 20\,kyr and the bottom three panels are 25\,kyr, 30\,kyr, and 35\,kyr. }
\label{slice_z_fig}
\end{figure*}

In Figures~\ref{slice_y_fig} and~\ref{slice_z_fig} we show the density and velocity distributions of the model via slices through the $y$-axis and $z$-axis respectively. Radiation pressure starts to produce outflow cavities at around 10\,kyr, but because the disc is not yet fully established the outflow on one side of the disc dominates. The system because more symmetrical with time as two radiation-pressure dominated,  high velocity cavities emerge with a large, thick, accretion disc surrounding the protostar. The cuts through the $z$-axis show the emergence of a large (1500\,au radius) accretion disc, within which are large scale spiral structures driven by gravitational instabilities. 

\section{Observational comparisons}
\label{obs_sec}

How well does our model agree with observational constraints on massive star formation? We can make such comparisons at two levels. The first is to compare the gross characteristics (masses, accretion rates, luminosities) from our simulation with those found from observation. The second is to make a comparison at a level that is closer to the observations themselves, by synthesising how the model would look to an observer via so-called simulated observations. 

\subsection{Accretion rate}
\label{acc_rate_sec}

We start by comparing the accretion rate from the simulation with observations.  In a recent review of discs around luminous young stellar objects \cite{beltran2016} presented compilations of observational data of accretion rates as a function of stellar mass (their Figure~15).  They  explored four different estimators of the accretion rate: (i) The infall rate estimated from red-shifted absorption line profiles, (ii) by assuming that the infall velocity is equal to the rotational velocity, (iii) by assuming that all the material from the rotating structure around the protostar will accrete on a free-fall time, and (iv) by assuming that the accretion rate is one sixth of the outflow rate derived from observations of molecular outflows.

\begin{figure}
\includegraphics[width=80mm]{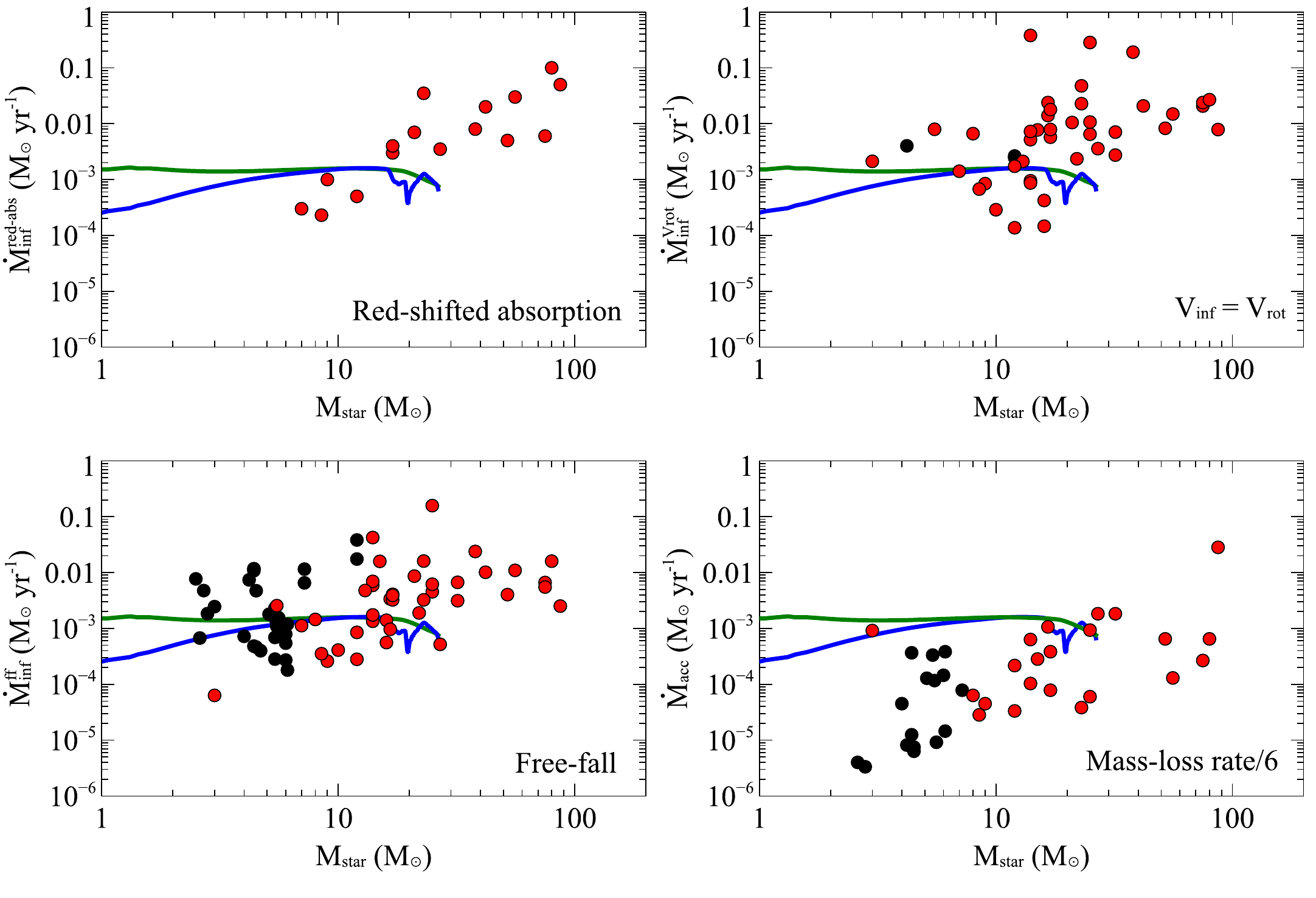}
\caption{A comparison between our model and observations in the mass-accretion rate--protostellar mass plane. The data are from  \protect\cite{beltran2016} and correspond to different estimates of the mass accretion rate detailed in Section~\protect\ref{acc_rate_sec}. The black and red circles correspond to intermediate- and high-mass protostars respectively. The blue line corresponds to the mass-flux through a 1500\,au spherical surface, while the green line represents the mass-accretion rate onto the protostar (smoothed over 100\,yr intervals).}
\label{beltran_fig}
\end{figure}

In Figure~\ref{beltran_fig} we show the evolutionary track of our model in the $\dot{M} - M_*$ plane against the data collated by \cite{beltran2016}. We show two different measures of $\dot{M}$ from our simulation, the first being the accretion rate onto the protostar (smoothed over 100\,yr intervals), and the second being the mass flux passing through a spherical surface of radius 1500\,au (the approximate radius of the protostellar disc, see below). The latter rate might be regarded as more representative of the observational rates, which probe large scales. The agreement with observation is generally good (although there is significant scatter in the data), but the model accretion rates seem to be systematically at the lower end of the distribution of observational rates (apart from those derived by scaling the outflow rates).

At the end of our simulation the bolometric luminosity of the system is $1.77 \times 10^5$\,L$_\odot$, of which the accretion luminosity contributes $3.7 \times 10^4$\,L$_\odot$ (giving a ratio $L_{\rm acc}/L_{\rm bol} = 0.2$). The values are in good agreement with the high mass sample presented in \cite{beltran2016}.

\subsection{Disc mass and radius}

We work on the basis that the disc in the numerical simulations corresponds to the volume within which the gas mass density exceeds $10^{-15}$\,g\,cc$^{-1}$. Although apparently arbitrary, this threshold matches a by-eye estimate of the disc extent from visualisations of the simulations (see Figure~\ref{slice_z_fig}), and we note that the same density threshold was employed by \cite{krumholz_2009} when examining the properties of their model disc. The disc is not completely circular at this density cut due to large spiral density waves, but varies between 1000 and  2000\,au in radius, and we adopt 1500\,au as the characteristic radius of the disc. We find the disc mass by summing the density of all cells within 1500\,au of the protostar that has a density exceeding $10^{-15}$\,g\,cc$^{-1}$ . By the end of the simulation the disc mass is about 7\,\msol, or about 30 percent of the protostellar mass. 

We compare our disc mass and radius with observationally derived qualities in  Table~\ref{disc_table}. Broadly speaking the gross properties of our model disc are comparable with observations, falling into approximately the middle of the distribution in terms of mass, radius, and disc-to-protostellar mass ratio. 
 
 \begin{figure}
\includegraphics[width=80mm]{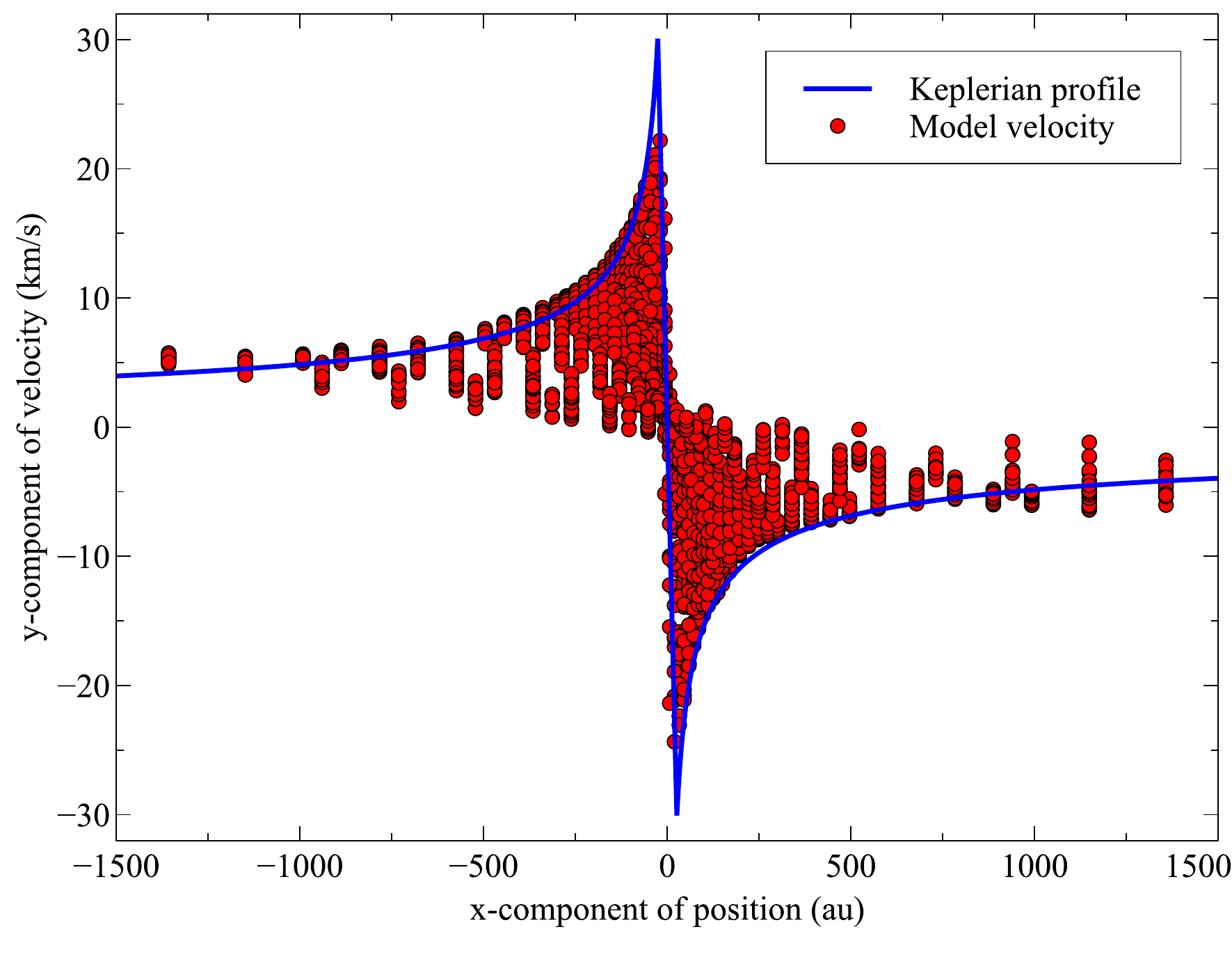}
\caption{The  $y$-component of velocity of cells in the mid-plane for the $t=35$\,kyr RHD model plotted against the $x$-component of the cell centre position (red circle). The Keplerian rotation curve for the sink particle mass (blue line) is also plotted.}
\label{keplerian_fig}
\end{figure}
 
We extracted the $y$-component of velocity of cells in our model disc which have densities greater than $10^{-15}$\,g\,cc$^{-1}$ and that have a face in the $z=0$ plane (i.e. the mid plane of the disc). We find that the disc is in Keplerian rotation (see Figure~\ref{keplerian_fig}). We note that  \cite{ilee2016}, \cite{johnston2015} and \cite{chen2016} identify Keplerian rotation in their observations of massive discs.  However the observational measurements of the disc kinematics are complicated by the fact that they are obtained from molecular line tracers that may also simultaneously probe the rotating and infalling envelopes (for example \cite{hunter2014} observed a combination of Keplerian rotation and infall). In order to make a more realistic comparison between the model and observations it is necessary to compute synthetic observations from the model and test the model more directly via molecular line profiles (see Section~\ref{molecular_sec}).
 
\begin{table*}
\caption{Observed properties of massive protostellar disc candidates along with the model disc parameters from this study. The columns give the object designation, the estimated mass of the disc, its estimated radius, the approximate disc-to-star mass ratio, and the reference.}
\label{disc_table}
\begin{center}
\begin{tabular}{lcccl}
\hline
Object & Mass (\msol) & Radius (au) & M$_{\rm disc}$/M$_*$ & Reference \\
\hline
NGC 6334 I(N), SMA 1b & 10--30 & 400  &  $\sim 0.5$ & \cite{hunter2014} \\
IRAS 16547--4247 & 6 & 500 & $\sim 0.5$ & \cite{zapata2015} \\
G11.92-0.61 MM1 & 2--3 & 1200 & $\sim 0.05$ & \cite{ilee2016} \\
AFGL 4176 & 12 & 2000 & $\sim 0.5$  & \cite{johnston2015} \\
IRAS 20126+4104 & 1.5 & 858  &$\sim 0.13$ & \cite{chen2016} \\
Our model & 7 & 1500 & 0.3 & This work \\
\hline
\end{tabular}
\end{center}
\end{table*}

\subsection{Spectral energy distributions}

\begin{figure}
\includegraphics[width=80mm]{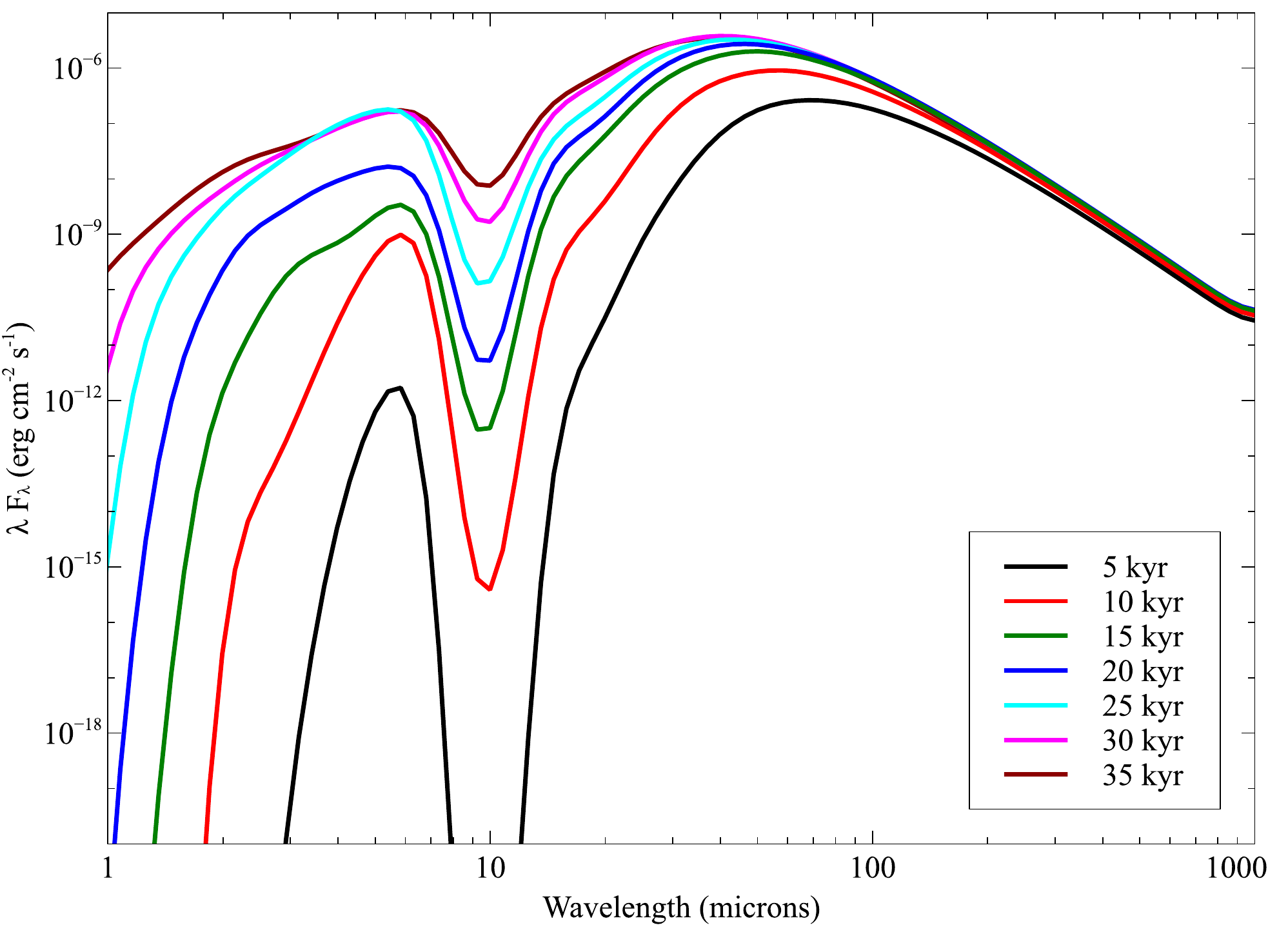}
\caption{Synthetic spectral energy distributions plotted logarithmically as $\lambda F_\lambda$ against wavelength in microns. We plot 7 snapshots of the model from computed at times ranging from 5\,kyr to 35\,kyr. The colour of the spectrum corresponds to the time of snapshot (given in the figure key). Note the strong 10\,\micron\ feature in absorption.}
\label{sed_fig}
\end{figure}

We calculated the spectral energy distributions (SEDs)  from the models density and temperature distributions assuming an inclination of 60$^\circ$, and these are plotted in Figure~\ref{sed_fig}.  Strong 10\,$\mu$m absorption features are seen in all spectra, consistent with its deeply embedded nature and the presence of small silicate grains. 
The rapid growth of the near- to mid-IR flux is seen as the model evolves and the central object becomes more luminous. 

\begin{figure}
\includegraphics[width=80mm]{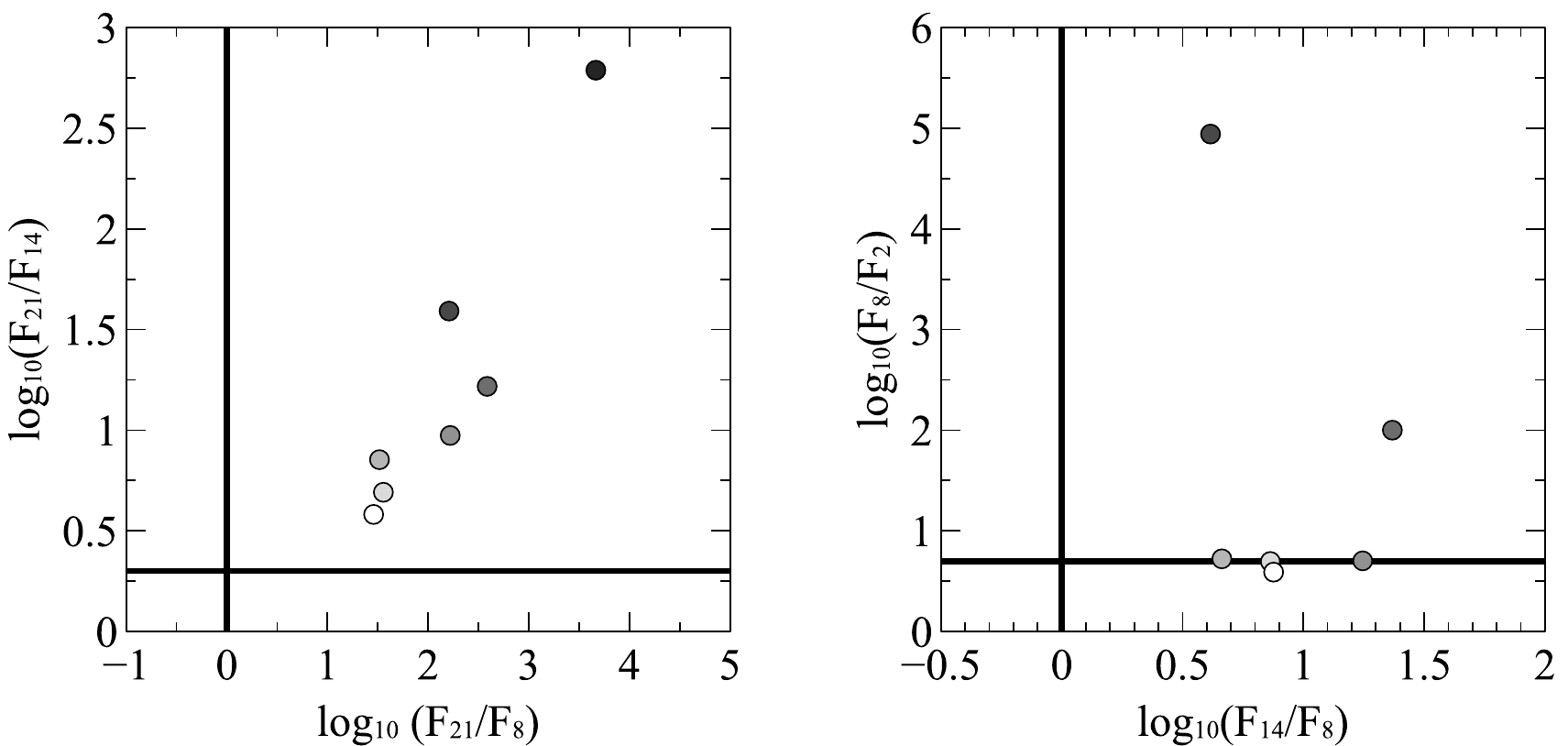}
\caption{Synthetic fluxes for the model (circular symbols) along with RMS survey cuts (thick solid lines). The symbol colours denote the time for the model, with black corresponding to 5\,kyr and white to 35\,kyr. The left-hand panels shows cuts $F_{21}/F_{8} > 2$ and $F_{21} > F_{14}$. The right-hand panel shows $F_{14} > F_{8}$ and $F_8 / F_K > 5$. (Note that $F_8 / F_K$ for the 0\,kyr model is not plotted, but has a value $>20$.)}
\label{cuts_fig}
\end{figure}

Since we know that the model is meant to represent a MYSO it is worth testing whether our object would appear in the Red MSX Survey catalogue \citep{lumsden2013}, which collates data on a large number of MYSOs selected from the point source catalogue of the MSX satellite \citep{price_2001}. In order to do this we constructed synthetic fluxes in the appropriate wavebands assuming a canonical distance of 1\,kpc. The cuts used by \cite{lumsden2013} are $F_{21} > 2F_8$, $F_{21} > F_{14}$, $F_{14} > F_8$, $F_8 > 5F_K$ and $F_K > 2F_J$. (Note that since our model is still deeply embedded at the end of the simulation we neglect the last constraint.). We plot the colour cuts in Figure~\ref{cuts_fig}, and do indeed find that the first three cuts are easily passed at all times. However we note that the $F_8 > 5F_K$ constraint is {\em not} passed for the later models, and at we attribute this to a deficit of $F_8$ flux due to the presence of a strong silicate 10\,$\mu$m feature.

\subsection{Images}

Before we proceed to examine synthetic images of our models it is worth  taking a short digression into the jargon of the field of massive star formation, focussing on the term `bipolar outflow cavities'. At early times our simulations show  (see Figure~\ref{slice_y_fig}), a pair of roughly diametrically opposed rarified regions that are formed by a rapid ($\sim 100$\,\kms) quite collimated, outflows driven by radiation pressure. As the simulation progresses and the disc forms, the outflows straighten towards the rotational axis and the hot gas inside the rarified regions expands to form a pair of `bubbles'. However, bipolar cavities may also be carved by highly-collimated magnetically-driven outflows (jets), and these are typically characterised by much smaller opening angles. The self-consistent calculation of the formation of these magnetic towers requires full radiation-magneto-hydrodyamics at a resolution that would render large-scale simulations intractable, but some assessment of their role in modifying the effects of radiation pressure may be made by incorporating a sub-grid model of the bipolar outflow e.g. \cite{cunningham_2011} and \cite{federrath_2014}. In the following discussion we refer to bipolar cavities, and cavity walls, in reference to the rarified structures that are formed along the rotation axis of our simulations, and defer an investigation of magnetically-driven outflows to future work.

We calculated synthetic images of the model at 35\,kyr. These monochromatic images were calculated at the nominal wavelengths of the MSX passbands and also Herschel PACS 70\,\micron, Herschel SPIRE 250\,\micron\ and 350\,\micron, and ALMA 850\,\micron. The 2\,\micron\ image is dominated by protostellar light scattered in the walls of the bipolar cavities. The mid-IR images show significant thermal emission from the warm dust at the base of  cavities, and the 70\,\micron\ is produced by dust throughout the cavities and the cavity walls. The longer wavelength images are dominated by cool, dense dust in the disc around the protostar.

\begin{figure*}
\includegraphics[width=180mm]{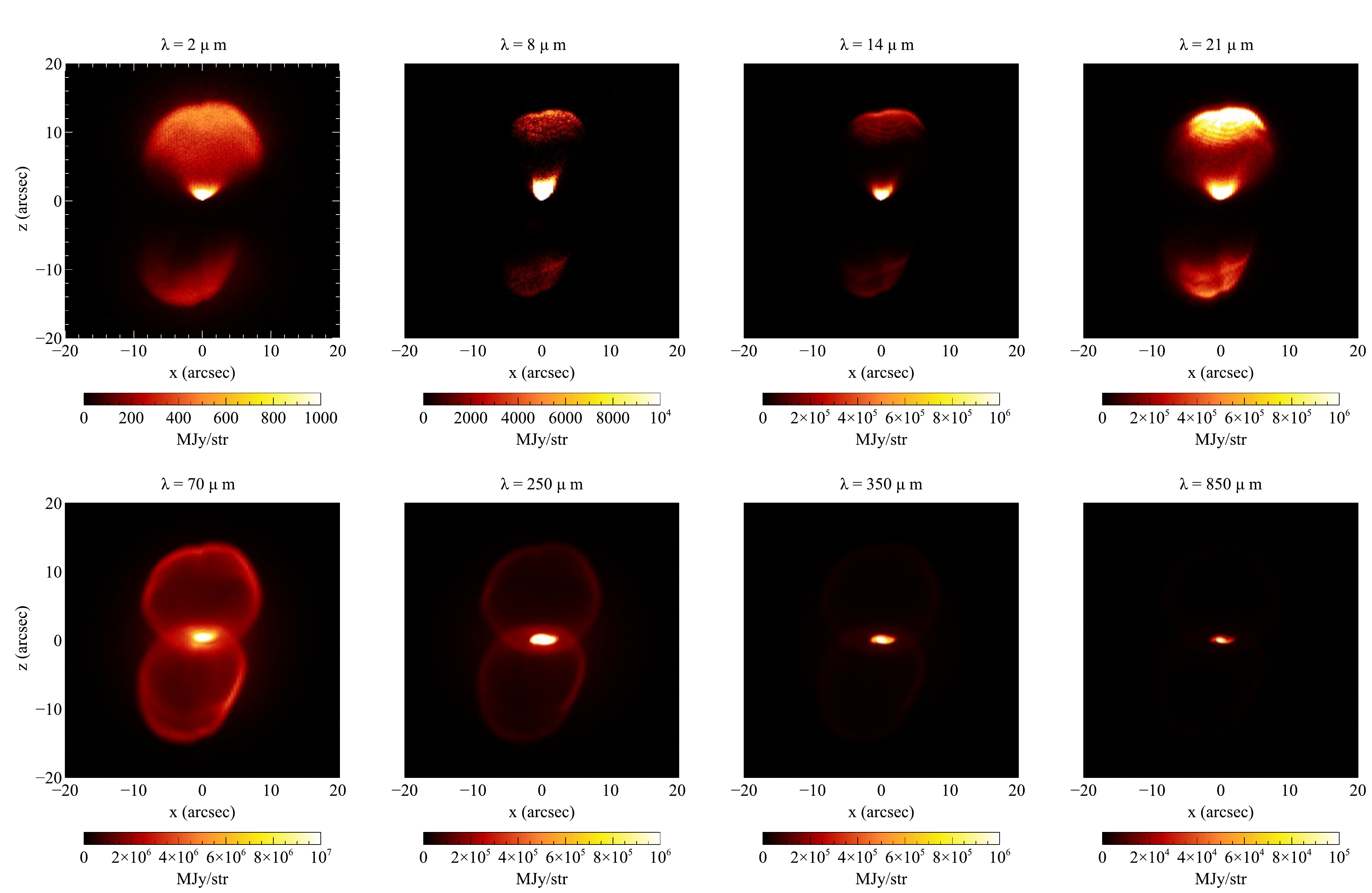}
\caption{Simulated images of the radiation-hydrodynamical model at 35\,kyr viewed at an inclination of 60$^\circ$ assuming a distance of 1\,kpc. The linear colour scales represents the surface brightness in MJy/str. Note that the maximum of the colour scale varies from image to image.}
\label{images_fig}
\end{figure*}

\cite{alvarez_2004} presented  H and K band speckle imaging of outflows in MYSOs. They found that 6 objects from their sample of 21 showed a conical nebula which could be interpreted as arising from scattered light from the dusty walls of the outflow cavity. Typically these nebulae had an extent of about half an arc second, and are monopolar in nature, the implication being that scattering from the more distant cavity is obscured by intervening material such as a disc. Some of the objects were found to have a gap between the location of the protostar and the base of the nebular cone. Our 2\,\micron\ image (top left of Figure~\ref{images_fig}) is consistent with the \cite{alvarez_2004} observations, although we note at 60$^\circ$ inclination the image peak occurs at the position of the protostar since there is substantial unscattered near-IR thermal emission from the  unresolved inner disc and protostar. However, although there is some low surface brightness emission at large angular distances from the protostar, the image is dominated by a conical, monopolar nebula whose centre of brightness is offset from the location of the protostar. The angular extent of the conical nebula (about 2 arc seconds) is rather larger than the typical size of those observed by \cite{alvarez_2004} at our 1\,kpc distance. 

\cite{dewit_2009} obtained diffraction limited (0.6 arc second) 24.5\,\micron\ images of 14 massive star formation regions. These images revealed the presence of MYSO sources that are resolved on arc second scales, and that they are (to first order) circular on the sky. They found that it was possible to simultaneously fit the imaging and SED data for the objects using spherical models with a $p=1$ density profile, although there were a subset of objects which were not well-fitted by spherical models. They conclude that these objects are viewed either edge-on or face-on. Our 21\,\micron\ image (top right of Figure~\ref{images_fig}) is dominated by thermal emission from the  disc and the warm cavity walls. We note that the peak surface brightness of our model (about 10$^7$\,MJy/str) is similar to those found by \cite{dewit_2009} but clearly the image is not circularly symmetric on the sky. 

\cite{olguin_2015} found that three MYSOs at distances between 4--5\,pc were marginally resolved at 70\,\micron\ in Herschel Hi-GAL survey data, and also found that the data could not be adequately fit with spherical models. Using two-dimensional RT models it was found  that a lower power-law density index (0.5) than that of pure infall (1.5) was necessary to fit the data, and they attributed this to extended emission from warm cavity walls. We re-scaled our fiducial 70\,\micron\ image to 5\,kpc distance and smoothed it to Herschel PACS resolution. We find that our model image would be very marginally resolved at this distance and resolution, but that an observer might be able to discern that the object was elongated.

\begin{figure}
\includegraphics[width=80mm]{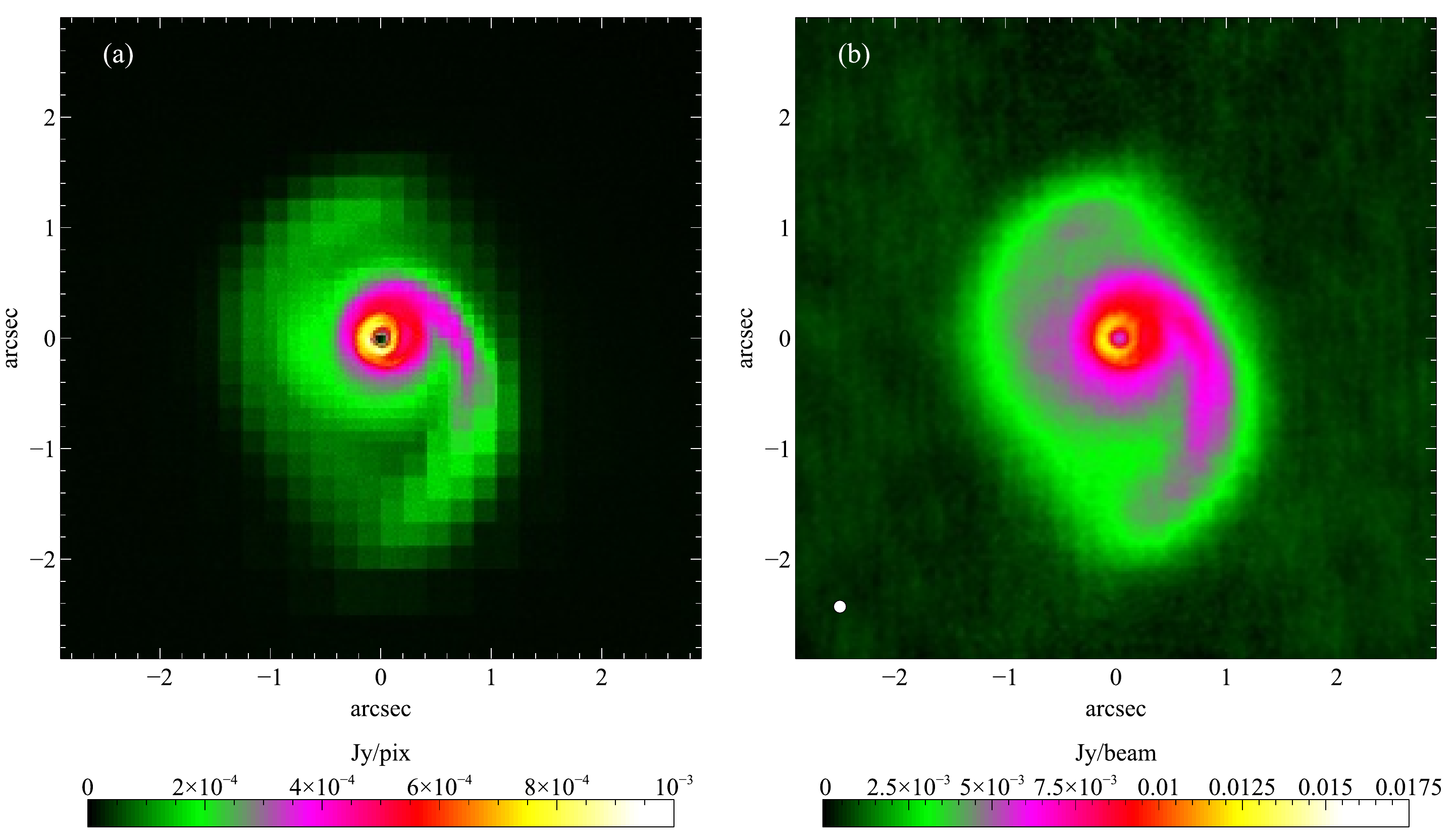}
\caption{A 1\,mm thermal dust image of the model at 35\,kyr viewed pole-on (left panel). The result of the CASA simulation is shown for comparison (right panel) along with the size of the beam (filled white circle).}
\label{alma_fig}
\end{figure}

As mentioned in Section~\ref{model_sec} the disc shows significant spiral structure. In order to ascertain whether this predicted structure could be observed by the Atacama Large Millimetre Array (ALMA) we constructed a pole-on monochromatic thermal dust continuum image at a wavelength of 1\,mm using the 35\,kyr model. This was then used as the basis  of a synthetic ALMA observation using CASA \citep{mcmullin2007}. For this image baselines were combined from extended and intermediate ALMA configurations (finished ALMA configurations 6 and 20, as included in the CASA software package) and the total time used was 80 minutes (1 hour for the extended configuration and 20 minutes for the intermediate). Thermal noise was added to the image using an assumed amount of precipitable water vapour of 0.5\,mm. The resulting image was cleaned interactively down to the level of 0.2\,mJy/beam. The resulting synthetic ALMA images is shown in Figure~\ref{alma_fig} (right panel) with the radiative-transfer simulation used as the skymodel for comparison (left panel). One can see that the large-scale structure observed in the radiative-transfer model is beautifully recovered in the CASA simulated ALMA observation. Furthermore the sublimation radius (seen as the black spot in the centre of the radiative-transfer model) is also observed in the CASA simulation as a dip in flux in the centre of the model.

\subsection{Molecular tracers}
\label{molecular_sec}

Many of the recent advances on the dynamics of discs around MYSOs have been dependent on high resolution molecular observations. We have not treated time-dependent chemistry during our simulation, but by assuming a simple prescription for the abundance distribution of the target molecule we can use the molecular line module of {\sc torus} to simulate the line emission data cube based on the RHD model's density and temperature distribution. 

We focus on transitions of the methyl cyanide molecule (acetonitrile, CH$_3$CN) since the high excitations transitions of the $k$-ladder have been used to probe massive star discs.  We adopt the same abundance distribution relative to H$_2$ as \cite{johnston2015} i.e. 
\[
X_{\rm CH_3CN} = \left\{ \begin{array}{ll} 
10^{-8}  & T > 100\,{\rm K} \\
5 \times 10^{-9} & 90\,{\rm K} < T < 100\,{\rm K} \\
10^{-10} & T < 90\,{\rm K} 
\end{array}
\right. \]
 
Using this prescription we found that the emission from the system had a significant component from the hot walls of the bipolar cavities and very strong emission near the sink particle, which is not something that is observed. We therefore added the further constraint that only gas with a mass density of greater than $10^{-15}$\,g\,cc$^{-1}$ and a gas temperature of less than 1500\,K had a non-zero CH$_3$CN abundance.
 
The molecular line calculations were computed with {\sc torus}, using the molecular RT physics module as described and tested in \cite{rundle_2010}. We used the CH$_3$CN energy levels from the {\sc lambda} database \citep{schoier2005} and we took the collisional rates from \cite{green1986}. We focus on the  $J=13 \rightarrow 12$, $K=3$ (239.096411 GHz) transition, which has a critical density of $4 \times 10^6$\,cm$^{-3}$ at 500\,K. The gas density in the protostellar disc exceeds $10^8$\,cm$^{-3}$ and we therefore assumed local thermodynamic equilibrium, adopting the gas temperatures from the RHD calculation. Using the $t=35$\,kyr RHD snapshot we computed a data cube with $256 \times 256$ spatial pixels (corresponding to $6.2$ arc seconds on a side at an assumed distance of 1\,kpc) and 33 velocity channels ranging from $-8$\,\kms\ to $+8$\,\kms. An inclination of 30$^\circ$ was adopted.

\begin{figure}
\includegraphics[width=80mm]{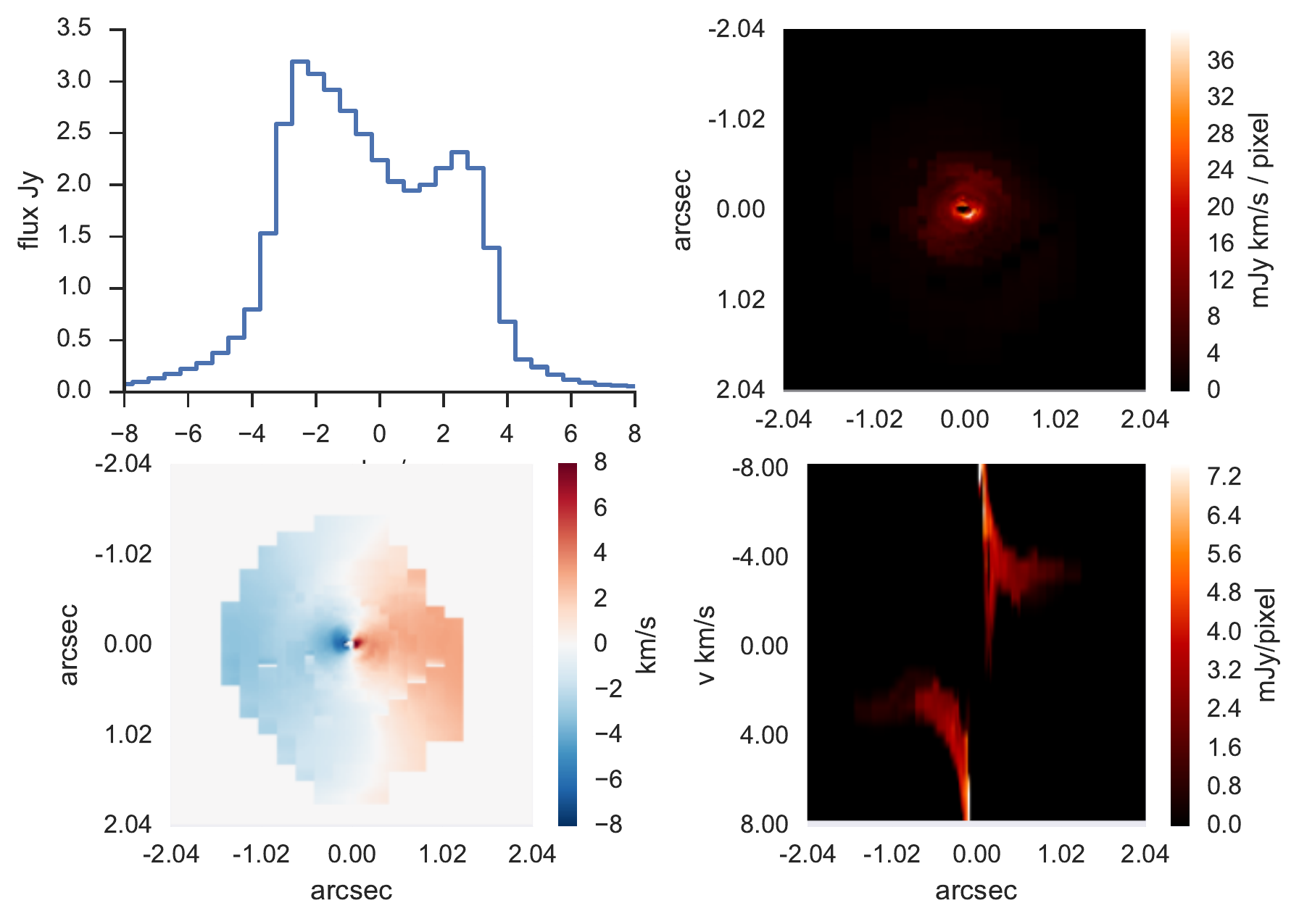}
\caption{The line profile (upper left), integrated intensity map (upper right in Jy\,\kms/pixel), moment 1 map (lower left) and position velocity diagram (lower right in mJy/pixel), for the CH$_3$CN $J=13 \rightarrow 12$, $K=3$ line. }
\label{ch3cn_fig}
\end{figure}

Figure~\ref{ch3cn_fig} shows the total integrated spectrum of the $J=13 \rightarrow 12$, $K=3$  transition (upper left). A double-peaked profile is seen, with a significantly stronger blue peak due to the asymmetric nature of the disc. The brighter blue peak is highly dependent upon the viewing angle to the model however, a different choice of viewing angle can result in a stronger red peak or two equally bright peaks. The continuum-subracted zeroth order moment map (upper right) shows that the emission is strongly peaked towards the centre of the disc, as expected for this high excitation line. The integrated intensity map also shows a gap in emission at the very centre of the disc where CH$_3$CN is depleted due to the high temperature. The first-order moment map (lower left) shows the expected pattern for a Keplerian disc. The position velocity diagram (lower right) shows that the rotation curve of the disc could be extracted from these CH$_3$CN observations if they were made with sufficient resolution and sensitivity.  We note that similar calculations were performed by \cite{krumholz2007}, and although they assumed a constant CH$_3$CN abundance, they drew broadly the same conclusions regarding the feasibility of extracting kinematic information from the observations.

Four of the five observations listed in Table~\ref{disc_table} (the observation of IRAS 16547 does not include CH$_3$CN as one of its targets) included measurements of a methyl cyanide (CH$_3$CN) line, and in each of them the velocity extent is 10--15 km$\,$s$^{-1}$ across the line. This is in agreement with the $\sim$10 km$\,$s$^{-1}$ seen in our synthetic observations of the CH$_3$CN $K = 3, J = 13 \rightarrow 12$ line.  The  peak surface brightness in this transition is slightly stronger (a factor of 1.5--2) than the observations of \cite{chen2016} and \cite{ilee2016}.

\section{Discussion}
\label{discussion_sec}

Our simulation is consistent with a picture of massive star formation as a scaled-up version of the low-mass paradigm, with stochastic disc accretion onto a protostar from a Keplerian disc accompanied by bipolar outflows. Since we have adopted a novel radiative-transfer scheme for this calculation it is interesting to compare our results to other recent studies.  \cite{klassen_2016} studied high mass star formation by  implementing  the hybrid radiation transport scheme in the {\sc flash} AMR hydrodynamics code. They started with spherical protostellar cores of 30\msol, 100\msol, and 200\msol, with an $\rho(r) \propto r^{-3/2}$ density profile and the same solid body rotation as this study. Broadly the result of their 100\msol\ simulation is very similar to our calculation, with the presence of a thick disc showing evidence of spiral structures and fast, bipolar cavities that initially show significant asymmetries (these authors also attribute these asymmetries to resolution effects due to the carteisan mesh). The accretion rate onto the star is similar to ours during the early phases of the calculation, and both show the accretion rate initially peaking after 5--10\,kyr. However the \cite{klassen_2016} shows a rapid growth in the accretion rate as the disc becomes gravitationally unstable. We attribute the difference to the fact we have fixed our protostar at the origin for the purpose of this first calculation, which will suppress gravitational instabilities that would otherwise enhance accretion. Despite this difference the similarity between the density slices in Figure~\ref{slice_y_fig} and those of Figure~13 in \cite{klassen_2016} is remarkable.

There has been considerable controversy in the literature regarding the role of radiatively-driven Rayleigh-Taylor (RRT) instabilities during the formation of massive stars. First suggested by \cite{krumholz_2009} RRT instabilities arise from when the dense gas in the cavity shells starts to penetrate the rarified gas of the cavities via finger-like intrusions. Numerical simulations focussing on galactic scales confirm the presence of such instabilities in optically-thick radiation-pressure driven dusty winds \citep{krumholz2012,davis2014, rosdahl2015,tsang2015}, although the details of the dynamics are found to be sensitive to the adopted radiation treatment.

\cite{kuiper_2012} argued that the RRT instabilities were a result of the underestimate of the radiative driving due to the use of grey flux-limited diffusion by \cite{krumholz_2009}, whereas the hybrid method they employed better captured the momentum deposited by the radiation at the $\tau=1$ surface. It appeared that the results of \cite{klassen_2016}, who also used a hybrid method, supported this finding, since they did not observe RT instabilities during their simulations. Recently \cite{rosen_2016} returned to this issue and implemented a hybrid method into the same hydrodynamics code used by \cite{krumholz_2009}. They found that although the implementation of the hybrid scheme suppresses the onset of the RT instabilities they nonetheless still occur. They argue that since the star is fixed in the Kuiper et al. simulations the optical depth effects that can stochastically occult the radiation field and suppress the radiation driving are absent, leading to a more stable outflow. They also argue that  the Kuiper calculations (which are performed on a fixed spherical mesh) do not have a sufficient resolution at the cavity boundary ($\sim 400$\,au) to capture the growth of the small scale perturbations that should seed the instability. They point out that although the Klassen simulations use an adaptive mesh they do not refine on the gradient of the radiation density, and thus the resolution along the cavity ($\sim 160$\,au) is insufficient to resolve the instability. Note that for comparison the calculation presented here has a resolution of 418\,au at a radius from the protostar of 4000\,au (roughly the radius at which the instabilities occur in the Rosen et al. simulation).

The question seems to be not whether RRT instabilities can occur, but whether a particular simulation is conducted in a regime where the instability is important. It seems likely that in some published simulations the radiation pressure of the direct (protostellar) radiation field at the dust sublimation radius imparts sufficient momentum that the outflow  cannot be halted by infalling material in the outer envelope, and reprocessed radiation plays no role in the driving process. Such a simulation would not be expected to show RRT instabilities. However for different initial conditions the momentum imparted to the gas at dust sublimation radius may be insufficient to drive a fast, stable outflow and instead the reprocessed radiation field is an important driving mechanism, and RRT instabilities result. Of course the preliminary simulation presented here neither has neither a  moving star nor a grid that refines on the radiation field, so although we find good agreement with the \cite{klassen_2016} simulations we defer a detailed discussion of the role of RT instabilities to future work. 

Our RHD model finishes with a luminous protostar, surrounded by a thick disc, large  cavities, and an optically think infalling envelope. The bolometric luminosity, and the fraction of accretion luminosity to protostellar luminosity are in good agreement with observations. The model SEDS (see Figure~\ref{sed_fig}) are comparable to those in the RMS survey (see, for example, \citealt{mottram_2011}).  Of course for these deeply embedded objects much of the leverage of the SED on the density structure is lost as the radiation field is reprocessed by the optically thick envelope prior to being observed. 

Extending the comparison to imaging we predict near-IR monopolar conical nebula caused protostellar and inner-disc emission scattering  off the cavity walls. The extent of the scattered light nebula appears rather larger than the observations, although the morphology of the nebula is  sensitive to the adopted viewing angle. The presence of the bipolar cavities also has a strong impact on the mid-IR images, with the 21\,\micron\ image particularly showing extended emission from the cavity walls. The fact that 24.5\,\micron\ imaging shows mainly circular symmetric emission places a strong constraint on our models, and suggests that our outflow cavities are overly large in the RHD model.  

It is interesting to compare our synthetic images with those from analytical prescriptions of the disc and envelope density structure published in a sequence of papers by Zhang and co-workers \citep{zhang_2011, zhang_2013, zhang_2014}. The outflow cavity in these models is assumed to arise from a disc wind which evolves in such a way that the cavity opening angle increases with time, reaching approximately 30$^\circ$ when the protostar is 10\,\msol\ and extending to 90$^\circ$ when the protostar is 24\,\msol. Unfortunately the analytical density distribution for the 24\,\msol\ model in Zhang et al. (2014; hereinafter Z14)  is sufficiently different to our RHD simulation that the images do not compare well. However, the 16\,\msol\ model images (Figure 21 of Z14) do show some obvious qualitative similarities with our synthetic images (Figure~\ref{images_fig}). In particular the 20\,\micron\ image show is significantly extended in Z14, in agreement with our models and in contradiction with observations. The 2\,\micron\ scattered-light image in Z14 is even more extended than our own, and is therefore also large compared to observations.

The root cause of this disagreement is probably the lack of collimated magnetically-driven jets in our calculations (and those of Z14), which may result in bipolar cavities with smaller opening angles at later times. We are investigating this using two-dimensional simulations and the sub-grid jet model of \cite{federrath_2014}, and will report on these simulations in a future paper.

Fortunately as we go to longer wavelengths the agreement with observation improves, and we predict that our system would only be marginally resolved at 70\,\micron\ by Herschel PACS, in agreement with the findings of \cite{olguin_2015}. Processing a 1\,mm dust continuum image with the CASA software suggests that ALMA should be able to resolve the spiral waves that occur in our disc, and even perhaps the dust sublimation radius.

Our synthetic molecular line observations suggest that kinematical data should be straightforwardly extracted (see Figure~\ref{ch3cn_fig}), allowing a straightforward determination of the protostellar mass by assuming the disc is in Keplerian rotation. This is a  good assumption for our RHD model (see Figure~\ref{keplerian_fig}), but the observations paint a more complex picture, in which the infalling envelope has an effect on the molecular line RT.  Clearly the main limitation is the lack of a chemistry model to predict the abundance distribution of CH$_3$CN. Although it is possible to couple RHD with time-dependent chemistry for limited reaction networks it would currently be intractable to attempt such a computation of a network detailed enough to follow the formation of CH$_3$CN. However an equilibrium chemistry post-processing of the RHD model is currently within our reach, and we are working on implementing this with the {\sc torus} framework.

\section{Conclusions}
\label{conc_sec}

We have presented a new simulation of the formation of a massive star
using a novel MC-based RHD method. Our model shares characteristics
with RHD simulations based on hybrid RT methods
(e.g. \citealt{kuiper_2013}, \citealt{klassen_2016}), such as a large,
thick protostellar disc and fast, broad, outflow cavities. The bulk of
the protostellar mass is acquired via stochastic disc accretion. 

We do not observe radiatively-driven Rayleigh-Taylor instabilties, but
as noted above we do not refine our AMR mesh at the cavity shell,
which may mean that we may under-resolve the small-scale variations that
seed the instability \citep{rosen_2016}. However we are currently in the process of
running a suite of 2D (cylindrical) models and we find that at early
times the bipolar outflows may be overwhelmed by infalling
material. This is primarily due to variability in the protostellar
accretion rate changing the driving luminosity, but it sensitive to
initial conditions. In order to fully resolve this issue we are
implementing a grid refinement method that refines the AMR method at
the cavity boundary.

We have attempted to compare our model with observations. We find that our model is consistent with observations in terms of protostellar luminosity and accretion rates, and that the protostellar disc matches the observed distribution in terms of disc mass and radius. We predict the disc to be in Keplerian rotation, but with significant large-scale gravitationally induced spiral structure that would resolvable with ALMA. 

The SEDs predicted by our models show good agreement with observed MYSO spectra, and that the synthetic colours of the object would mean that it would be selected as an MYSO using most of the colour cuts adopted for the RMS survey. However, we note that the strength of the 10\,\micron\ silicate feature may be sufficient to suppress the 8\,\micron\ flux in such a way that it might fail the $F_8 > 5F_K$ cut.

Synthetic images of RHD model show some features that are observed, such as conical scattered light nebula in the near-IR and 70\,\micron\ emission that would only be marginally resolved in Herschel PACS data. Although the peak surface brightness of our synthetic 21\,\micron\ imaging matches the observations well, its angular extent  does not, and this perhaps indicates that our RHD model is over-predicting the extent of the bipolar cavities. 

We show that synthetic molecular line observations of CH$_3$CN predicts the correct surface brightness in the line, as well as the line velocity width. We find a blue-red asymmetry in the line profile, and attribute this to asymmetries in the disc itself (as opposed to RT effects as the line emission from the disc passes through the infalling envelope). We note that our molecular line calculations are limited by a lack of self-consistent chemistry.

In summary, our RHD model shows similar characteristics to other contemporary studies in terms of the disc, envelope and outflow properties. We have shown that these properties are broadly consistent with observations across a wide variety of observables. In future studies will address the issues of radiatively-driven Rayleigh Taylor instabilities, and study the disc fragmentation and binary formation. We will address the effect of including a sub-grid model of the collimated magnetically-driven outflow on the accretion process. We also hope to couple the {\sc torus} code with a chemical model in order to make more self-consistent predictions of molecular line diagnostics.

\section*{Acknowledgments}

The calculations for this paper were performed on the University of Exeter Supercomputer, a DiRAC Facility jointly funded by STFC, the Large Facilities Capital fund of BIS, and the University of Exeter, and on the Complexity DiRAC Facility jointly funded by STFC and the Large Facilities Capital Fund of BIS. TJH and TAD acknowledge funding from Exeter's STFC Consolidated Grant (ST/M00127X/1). We thank Takashi Hosokawa for kindly providing us with the protostellar evolutionary model. We are grateful to Maite Beltran for providing the data for Figure~\ref{beltran_fig}, and we thank Dave Acreman, John Ilee and Tom Haworth for useful discussions. We thank the anonymous referee for a helpful report.

\bsp

\label{lastpage}
\bibliographystyle{mn2e}
\bibliography{torus}
\end{document}